\documentclass[twocolumn,floatfix,preprintnumbers]{revtex4-1}

\usepackage[utf8]{inputenc}
\usepackage[colorlinks=true,citecolor=blue,linkcolor=blue]{hyperref}
\usepackage[normalem]{ulem}
\usepackage{url}
\usepackage{graphicx,wrapfig,float,slashed,cancel}
\usepackage{amsmath,amssymb,epsfig,graphicx,xcolor,stmaryrd}
\usepackage{bm}
\usepackage{enumitem}
\usepackage{physics}

\definecolor{darkblue}{RGB}{1, 90, 173}

\newcommand{\UV}{\text{UV}}
\newcommand{\IR}{\text{IR}}
\newcommand{\mkk}{M_\text{KK}}

\newcommand{\bl}{\bar{\lambda}}
\newcommand{\muv}{m_\text{UV}}
\newcommand{\mir}{m_\text{IR}}
\newcommand{\lir}{\lambda_\text{IR}}

\newcommand{\vp}{\varphi}

\begin{document}


\title{A warped scalar portal to fermionic dark matter}

\author{Adrian Carmona}
\email{adrian@ugr.es}
\affiliation{CAFPE and Departamento de F\'isica Teórica y del Cosmos, 
Universidad de Granada, E-18071 Granada, Spain}

\author{Javier Castellano Ruiz}
\email{castella@uni-mainz.de}
\affiliation{PRISMA$^+$ Cluster of Excellence {\em\&} Mainz Institute for Theoretical Physics, 
Johannes Gutenberg University, 55099 Mainz, Germany}

\author{Matthias Neubert}
\email{matthias.neubert@uni-mainz.de}
\affiliation{PRISMA$^+$ Cluster of Excellence {\em\&} Mainz Institute for Theoretical Physics, Johannes Gutenberg University, 55099 Mainz, Germany\\
Department of Physics {\em\&} LEPP, Cornell University, Ithaca, NY 14853, U.S.A.\\
Department of Physics, Universit\"at Z\"urich, Winterthurerstrasse 190, CH-8057 Z\"urich, Switzerland}

\date{\today}

\preprint{MITP/20-069}

\begin{abstract}
	We argue that extensions of the SM with a warped extra dimension, together with a new $\mathbb{Z}_2$-odd scalar singlet, provide a natural explanation not only for the hierarchy problem but also for the nature of fermion bulk masses and the  observed dark matter relic abundance. In particular, the Kaluza-Klein excitations of the new scalar particle, which is required to naturally obtain fermion bulk masses through Yukawa-like interactions, can be the leading portal to  any fermion propagating into the bulk of the extra dimension and playing the role of dark matter. Moreover, such scalar excitations will necessarily mix with the Higgs boson, leading to modifications of the Higgs couplings and branching ratios, and allowing the Higgs to mediate the coannihilation of the fermionic dark matter. We study these effects and explore the viability of fermionic dark matter in the presence of these new heavy scalar mediators both in the usual freeze-out scenario and in the case where the freeze-out happens during an early period of matter domination.

\end{abstract}


\maketitle

\section{Introduction} \label{sec:intro}

The discovery of the Higgs boson at the LHC represented the last step towards establishing the Standard Model (SM) as a solid theory describing the constituents of matter and their interactions down to very short distances. However, there are still some questions which do not have an answer within the SM. One of the most significant examples is the so-called hierarchy problem, the question why the Higgs boson is much lighter than the characteristic scale of gravity. One could argue that this problem is merely a by-product of our theoretical prejudices and that nature did not ask for a dynamical explanation for this difference of scales. Still, even in this case, we know for a fact that the SM cannot accommodate some other observed phenomena. One of the most striking examples is the existence of dark matter (DM). We know that there is no viable DM candidate in the SM, so already this fact asks for the presence of new physics. Extensions of the SM with a warped extra dimension (WED) compactified on an $S_1/\mathbb{Z}_2$ orbifold contain the necessary features for addressing simultaneously both of these issues, the absence of a viable DM candidate and the hierarchy problem. Moreover, they can also explain the large hierarchy existing between the different fermions masses, providing a calculable version of partial compositeness~\cite{Maldacena:1997re, Gubser:1998bc, ArkaniHamed:2000ds, Rattazzi:2000hs, PerezVictoria:2001pa, Contino:2004vy, Gherghetta:2010cj}, which makes them very attractive extensions of the SM.

Fermion masses in WED compactified on a $S_1/\mathbb{Z}_2$ orbifold need to have a dynamical origin, since the five-dimensional (5D) bulk masses must be $\mathbb{Z}_2$-odd functions on the orbifold \cite{Grossman:1999ra,Gherghetta:2000qt}. Indeed, one can easily see that the 5D Dirac fermion bilinear $\bar \Psi_i \Psi_i$ is odd under the orbifold symmetry, excluding the presence of a constant mass term. The most natural solution to this problem is to dynamically generate these masses with the help of a $\mathbb{Z}_2$-odd bulk scalar field. Indeed, if such scalar field develops a vacuum expectation value (VEV) with a non-trivial profile along the extra dimension, fermion bulk masses can arise dynamically through Yukawa-like interactions. We have explored this possibility in detail in \cite{Ahmed:2019zxm}, studying in particular its phenomenological consequences. In particular, one finds that the VEV has a kink-like profile, approaching the traditional sign function for large values of the odd scalar mass, whenever the WED is significantly larger than its inverse curvature. 

A natural question which arises in scenarios addressing the origin of the fermion bulk masses concerns the possible interplay with a bulk Higgs boson. In~\cite{Ahmed:2019zxm} we have considered a brane-localized Higgs field, which does not mix at tree-level with the $\mathbb{Z}_2$-odd scalar field. However, in the more general case of a Higgs boson in the bulk of the extra dimension~\cite{Davoudiasl:2005uu, Cacciapaglia:2006mz, Azatov:2009na, Vecchi:2010em, Archer:2012qa, Malm:2013jia, Archer:2014jca}, such a mixing is unavoidable and needs to be taken into account. Studying the effect of this mixing is one of the main goals of this work. On the other hand, since the odd scalar field is responsible for all fermion bulk masses, it represents a unique window into any femionic dark sector propagating into the bulk of the WED. Models with WEDs already feature an irreducible mediator between visible and dark sectors, since gravity couples to matter through the energy-momentum tensor. However, as we will see, when the DM candidates are fermionic weakly interacting particles (WIMPs) with masses of $\mathcal{O}$(TeV), the resonances arising from the 5D $\mathbb{Z}_2$-odd scalar field can provide the most important mediators for the DM coannihilation cross section. Moreover, due to the mixing with the bulk Higgs field, these fermionic dark sectors are mostly  Higgs-mediated for DM masses below the TeV scale. We examine thoroughly the resulting model of scalar-mediated fermionic DM for a large range of DM masses. We focus on the case where the DM particle is a vector-like (VL) fermion, the first Kaluza-Klein (KK) excitation of a 5D dark fermion. However, most of our results also hold in the case where the DM candidate gets an external mass, which can be chiral, VL or even of Majorana type.

This work is organized as follows: In order to set up the notation, we review the bulk Higgs case (disregarding the portal coupling) in section~\ref{sec:action}. In section~\ref{sec:mix} we solve the coupled system of field equations obtained after switching on the portal coupling between both types of bulk scalar fields by diagonalizing the resulting 4D mass matrix perturbatively. In section~\ref{sec:pheno} we proceed to discuss the phenomenology assuming a non-negligible portal coupling and the presence of $N_{\chi}$ dark fermion bulk fields. First, we discuss the impact of the scalar mixing on the SM Higgs couplings. We then continue by examining the impact of the dark fermions on the Higgs invisible decay width. Then, we discuss the predictions for the DM coannihilation cross-section mediated by the Higgs field and the first KK resonance of the $\mathbb{Z}_2$-odd scalar field, comparing these contributions with the ones mediated by KK gravitons. We compute the prediction for the  DM relic abundance as a function of the velocity-averaged coannihilation cross section in the usual freeze-out scenario as well as in the case of a matter-dominated universe~\cite{Gondolo:1990dk, Hamdan:2017psw, Chanda:2019xyl}. Finally, we compute the constraints arising from direct detection using data from the Xenon1T experiment, showing that for a $\mathcal{O}(10~$TeV) fermionic WIMP we can reproduce the observed DM relic abundance in the scenario of matter domination, without conflicting with current data from Xenon1T. In the case of radiation domination and DM masses of $\sim 15$~TeV, these scalar mediators can provide a non-negligible fraction of the required coannihilation cross section, even though additional mediators would be required. 

\section{A Bulk Higgs in a WED} \label{sec:action}

We consider a Randall-Sundrum (RS) model~\cite{Randall:1999ee} with the extra dimension compactified on an $S_1/\mathbb{Z}_2$ orbifold with two D3-branes localized at the fixed points, an ultraviolet (UV) brane at $t_\UV=\epsilon$ and an infrared (IR) brane at $t_\IR=1$, where $t$ is the coordinate describing the extra dimension. This coordinate is defined in terms of the usual $\phi = y/\pi$ coordinate by $t = \epsilon \, e^{kr \phi}$, where $\epsilon = e^{-kr\pi}\sim\mathcal{O}(10^{-16})$. In this notation, the metric reads
\begin{equation} 
	ds^2 = g_{MN}dx^Mdx^N=\dfrac{\epsilon^2}{t^2}\left( \eta_{\mu\nu} dx^\mu dx^\nu - \dfrac{dt^2}{\mkk^2} \right),	
\end{equation}
where $\mkk \equiv k \epsilon$ and $\eta_{\mu\nu}={\rm diag}(1,-1,-1,-1)$ is the 4D Minkowski metric. It is useful to define the quantity $L=kr\pi\sim 30$, which is a measure of the size of the extra dimension in natural units. Here $k$ and $r$ are the AdS curvature and the radius of the $S_1$. Note that in RS models addressing the hierarchy problem implies $L\gg 1$.


Let us start by reviewing the well-known case where the Higgs field does not mix at tree-level with the odd bulk scalar \cite{Archer:2014jca,Mahmoudi:2016aib}. For simplicity, a quartic term is only introduced on the IR brane in order to induce electroweak symmetry breaking (EWSB). The Higgs action reads
\begin{equation}
\begin{aligned}
	S = \int d^5x \sqrt{g}\bigg\{ & g^{M\!N} \left( D_{M}H \right)^\dagger D_{N}H-V(H) \\
	& - \sum_{k=\text{UV,IR}} \frac{\sqrt{|\hat g_k|}}{\sqrt{g}} \hat{V}^k(H) \, \delta(t-t_k)\bigg\}, \end{aligned}
\end{equation}
where $g=\det(g_{MN})$, $\hat{g}_k=\det(g_{\mu \nu}|_{t=t_k})$, and we define the Higgs doublet in the unitary gauge as
\begin{equation}
	H(x,t) = (0,\dfrac{t}{\epsilon\sqrt{2 r}} \left[ \vp_H(t) + h(x, t) \right])^T .
\label{eq:higgsfield}
\end{equation}
For the Higgs field and its VEV we follow the treatment of \cite{Malm:2013jia}. The bulk and brane-localized potentials for the bulk Higgs field are taken to be of the form
\begin{equation}
\begin{aligned}
& V(H) = \mu^2_H \abs{H}^2, \\
& \hat{V}^{\UV} = \sigma_{\UV} \abs{H}^2 , \\
& \hat{V}^{\IR} = - \sigma_{\IR} \abs{H}^2 + \rho_{\IR} \abs{H}^4 .
\label{eq:potential_bt}
\end{aligned}
\end{equation}
The equation of motion (EOM) for the Higgs VEV is
\begin{equation}
\begin{aligned}
\left[ t^2 \partial_t^2 + t \partial_t - \beta^2 \right] \dfrac{\vp_H}{t}= 0 , \qquad \beta^2 \equiv 4 + \mu_H^2/k^2 ,
\end{aligned}	
\end{equation}
along with boundary conditions (BC) on the UV and IR branes, which read
\begin{equation}
\begin{aligned}
& \partial_t \left[ t \, \vp_H(t) \right]_{t = \epsilon^+} = \muv \, \vp_H(\epsilon) , \\
& \partial_t \left[ t \, \vp_H(t) \right]_{t = 1^-} = \mir \, \vp_H(1) - \dfrac{2\lir}{\mkk^2} \vp_H(1)^3 . \label{eq:bchiggs}
\end{aligned}	
\end{equation}
The notation $\epsilon^+$ and $1^{-}$ refers to the orbifold fixed points, approached from the appropriate side. Above we have defined
\begin{equation}
	\muv = \dfrac{\sigma_{\text{UV}}}{2k} , \quad \mir = \dfrac{\sigma_{\text{IR}}}{2k} , \quad \lir = \dfrac{\rho_{\text{IR}}k}{4r} .
\end{equation}
This set of EOM and BCs leads to the well-known solution
\begin{equation}
 \vp_H(t) = N_v \left[ t^{1+\beta} - r_v t^{1-\beta} \right] , \label{eq:higgsvev1}
\end{equation}
where
\begin{equation}
\begin{aligned}
& r_v = \epsilon^{2\beta} \frac{ 2 + \beta - \muv }{ 2 - \beta - \muv}\, \,, \\
& N_v^2 = \frac{\mkk^2}{2\lir}\, \frac{\left(\mir - 2 - \beta \right) + r_v \left( \mir - 2 + \beta \right)}{(1 - r_v)^3} \,.
\end{aligned}
\end{equation}
Note that, unless $\beta$ is very small or $\muv$ is extremely fine-tuned to the value $2-\beta$, it is safe to set $r_v\propto \epsilon^{2\beta}\to 0$. Then, the Higgs VEV will be peaked towards the IR brane and expression \eqref{eq:higgsvev1} simplifies to 
\begin{equation}
\vp_H(t) \simeq N_v t^{1+\beta} = \vp_H(1) \, t^{1+\beta} ,
\end{equation}
with
\begin{equation}
 N_v \equiv \vp_H(1) \simeq \mkk \sqrt{\dfrac{\mir - 2 - \beta}{2 \lir}}. \label{eq:defNv}
\end{equation}
Demanding the normalization for the VEV to be such that one correctly reproduces the SM mass relations for the $W$ and $Z$ bosons leads to 
\begin{equation}
v_4^2 = \dfrac{2\pi}{L} \int^1_\epsilon \dfrac{dt}{t} \vp_H^2(t) = \dfrac{\pi}{L}\dfrac{\vp^2_H(1)}{1+\beta} + \mathcal{O}(\epsilon),	\label{eq:normSM}
\end{equation}
where we have used the fact that the zero-mode profiles of the gauge bosons are flat, up to corrections of $\mathcal{O} (v_4^2/\mkk^2)$, \cite{Malm:2013jia}. We can then write the VEV of the Higgs field as \begin{equation}\label{eq18}
\vp_H(t) \approx v_4 \sqrt{\dfrac{L}{\pi} (1 + \beta)} \, t^{1+\beta} ,
\end{equation}
where $v_4$ agrees with the SM parameter $v_{\rm SM}$ at leading order in $x_4 \equiv v_4/\mkk$.

\medskip

\section{Bulk Higgs and Odd scalar mixing}	\label{sec:mix}

We are now interested in the case where the two bulk scalar fields -- the Higgs and the new $\mathbb{Z}_2$-odd scalar -- mix with each other. In this case the action reads
\begin{equation}
\begin{aligned}
	S = \int d^5x \sqrt{g}\,\bigg\{ 
	& g^{M\!N} \sum_{i=1,2} \left( D_{M}\Phi_i \right)^\dagger D_{N}\Phi_i 
	 - V(\Phi_1,\Phi_2) \\
	& - \sum_{k=\text{UV,IR}} \frac{\sqrt{|\hat g_k|}}{\sqrt{g}} \hat{V}^k(\Phi_1,\Phi_2)\, 
	 \delta(t-t_k)\bigg\}, \label{eq:action}
\end{aligned}
\end{equation}
with $\Phi_1 = H$ and $\Phi_2 = (1/\sqrt{2}) \, \Sigma$ denoting the Higgs doublet and the (real) odd scalar field, which is a gauge singlet. We consider mixed BCs for the Higgs field, while the odd scalar field satisfies Dirichlet BCs. Such BCs for the bulk Higgs are a consequence of the brane-localized potentials, which are forbidden for the odd scalar, since it vanishes on the two branes. In our model, both bulk scalar fields develop a VEV. We can express the 5D odd scalar in terms of its background configuration $\varphi_S(t)$ and its 5D excitation $S(x,t)$ as 
\begin{equation}\label{Sigmadecomp}
	\Sigma(x,t) = \vp_S(t) + \dfrac{t}{\epsilon\sqrt{r}} \, S(x, t).
\end{equation}
The bulk potential reads 
\begin{equation}
	V(H,\Sigma) = \mu^2_H \abs{H}^2 - \frac{\mu_S^2}{2} \Sigma^2 + \frac{\lambda_S}{4} \Sigma^4 + \lambda_{HS}\abs{H}^2 \Sigma^2 ,	\label{eq:potential}
\end{equation}
where $\mu_H$, $\mu_S$ are the mass parameters and $\lambda_S$, $\lambda_{HS}$ the quartic couplings. The brane-localized potentials for the bulk Higgs field are the same as those shown in~\eqref{eq:potential_bt}. 

\subsection{Background solutions}

First, we determine the profiles of the two VEVs. From \eqref{eq:potential}, the EOMs are
\begin{widetext}
\begin{equation}
\begin{aligned}
\Bigg[ t^2 \partial_t^2 - 3 t \partial_t + \frac{\mu_S^2}{k^2} \Big( 1 - v_S^2 - \bl \dfrac{k^4}{\mu_S^4}\dfrac{\lambda_S}{r} t^2 \dfrac{\vp_H^2}{\mkk^2 } \Big) \Bigg] v_S(t) & = 0 , \\
\left[ t^2 \partial_t^2 + t \partial_t - \beta^2 - \bar{\lambda} v_S^2 \right] \dfrac{\vp_H(t)}{t} 
& = 0 , \label{eq:vescoupled1}
\end{aligned}	
\end{equation}
\end{widetext}
where we have defined the dimensionless coupling 
\begin{equation}
	\bar{\lambda} \equiv \dfrac{\mu_S^2}{k^2} \dfrac{\lambda_{HS}}{\lambda_S}
\end{equation}
and redefined the VEV of the odd field as in \cite{Ahmed:2019zxm}, i.e.\
\begin{equation}
	\vp_S(t) = \dfrac{ \mu_S}{\sqrt{\lambda_S}}\, v_S(t).
\end{equation}
In order to obtain an inverted one-dimensional Mexican-hat potential for $v_{S}$ and guarantee the existence of non-trivial solutions, we impose an upper bound on $\bl\lambda_S/r$. In practice, we demand that
\begin{equation}
\bl \dfrac{k^4}{\mu_S^4}\dfrac{\lambda_S}{r} t^2 \dfrac{\vp_H^2(t)}{\mkk^2 }\Bigg\vert_{t=1} \le 1.
\label{eq:max_condition}
\end{equation}
Since the Higgs VEV is monotonic in $t$ (at least at leading order in $\bar{\lambda}$), once this condition is fulfilled it will also hold for $t<1$. Plugging in the solution for the Higgs VEV for $\bar{\lambda}=0$ (i.e.\ for vanishing portal coupling $\lambda_{HS}$), this constraint translates into
\begin{equation}
\bl \dfrac{\lambda_S}{r} \lesssim \dfrac{x_4^{-2}}{10(1+\beta)} \left( \dfrac{\mu_S}{k} \right)^4,
\label{eq:max_condition_2}
\end{equation}
where $x_4=v_4/\mkk$.

\begin{figure}[t!]
\begin{center}
\hspace*{-0.2cm}
\includegraphics[scale = 0.5]{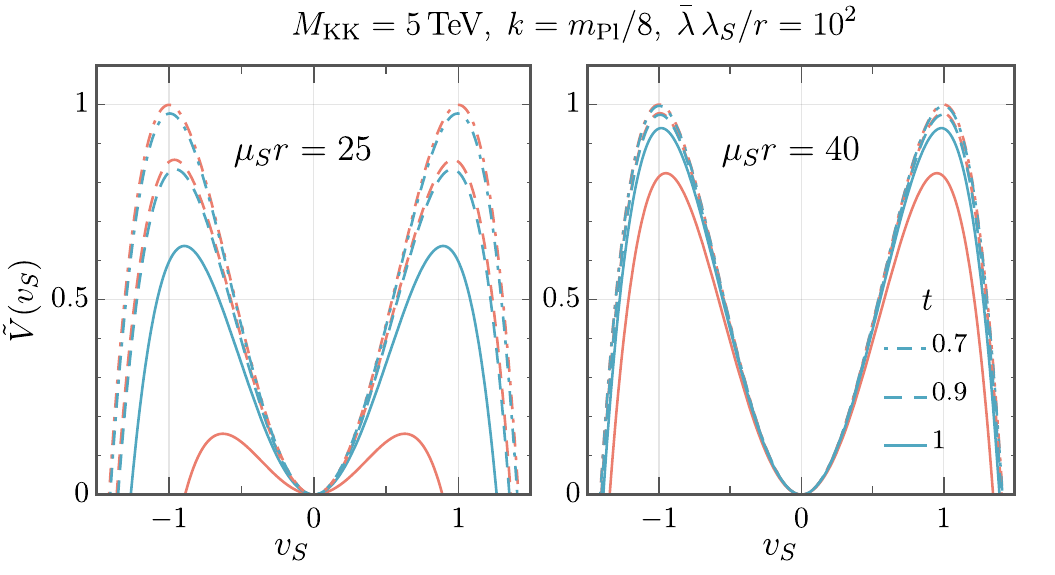}
\vspace*{-0.25cm}
	\caption{Effective potential $\tilde{V}(v_S)$ computed using  the leading-order solution $\vp_{H,0}(t)$, for different values of $t$ and  $\mkk=5\,$TeV, $k/m_{\rm Pl}=1/8$ and $\bar{\lambda} \lambda_S/r=100$, for $\beta=2$ (blue) and $\beta=8$ (red). } \label{fig:odd_pot}
\end{center}
\end{figure}

We can solve the coupled system of equations iteratively. The starting point are the solutions we already know for the decoupled equations -- $v_{S,0}(t)$ and $\vp_{H,0}(t)$. We thus insert the solution from the previous iteration for each VEV in the EOM corresponding to the other one in \eqref{eq:vescoupled1}. This method is convenient, since the potential for the odd VEV and the strategy to solve its EOM is well understood (see \cite{Ahmed:2019zxm}). Indeed, as we can see from figure~\ref{fig:odd_pot}, the potential does not differ much from the potential obtained in the decoupled case (zero portal coupling). We have checked numerically that the solution obtained for $v_S(t)$ fits the one obtained for the decoupled case with high accuracy. 

We display in figure~\ref{fig:odd_pot} the one-dimensional effective potential $\tilde{V}(v_S)$ defined by 
\begin{align}
	\frac{\delta \tilde{V}(v_S)}{\delta v_S}= \Big( 1 - v_S^2 - \bl \dfrac{k^4}{\mu_S^4}\dfrac{\lambda_S}{r} t^2 \dfrac{\vp_H^2(t)}{\mkk^2 } \Big)v_S ,
\end{align}
which determines the EOM for $v_S$ once $\varphi_H\approx \varphi_{H,0}$ is used, see (\ref{eq:vescoupled1}). We show this potential for $\bar{\lambda}\lambda_S/r=100$ and different values of $t$, $\mu_S$ and $\beta$, in order to explore how its maxima change with increasing $t$ and different values of $\mu_S$ and $\beta$. Note that in the figure we use the leading-order solution for the Higgs profile. Here and below we choose $M_{\rm KK}=5\,$TeV, which is motivated so as to avoid tensions with electroweak precision data (see~\cite{Ahmed:2019zxm} for more details). Similarly, we take $k=m_{\rm Pl}/8$ where $m_{\rm Pl} = 2.4 \cdot 10^{18}$ GeV is the reduced Planck mass. This value corresponds to $kr\approx 10.1$, or equivalently $\Lambda_{\pi}\equiv m_{\rm Pl}\,e^{-kr\pi}=40\,$TeV. We observe that both maxima decrease for increasing $t$. These two maxima would eventually collapse into the maximum of an inverted parabola, if values of $\bar{\lambda}\lambda_S/r$ larger than the ones given by equation \eqref{eq:max_condition_2} were considered. In that case, only the trivial solution would exist. In practice, we will never reach such large values due to perturbativity constraints on the first KK excitation of $S$, since the Yukawa couplings of this particle to the different fermions scale with $\sqrt{\lambda_S/r}$, see below. On the other hand, reproducing a value of the DM coannihilation cross section required to account for the observed relic abundance demands a large portal coupling between the different fermion sectors. For this reason, we consider $\lambda_S/r\lesssim \mathcal{O}(100)$  hereafter. In particular, in figure~\ref{fig:odd_pot} we choose $\bl\lambda_S/r=100$ to saturate the bound given by equation~\eqref{eq:max_condition_2}.

	\subsection{Scalar excitations}

We now move on to the study of the scalar KK excitations. The profiles of these resonances can be computed by inserting the KK decompositions
\begin{equation} 
\begin{aligned} 
h(x,t) & = \sum^\infty_{n=0} h_n(x) \chi_n^{h}(t), \\
S(x,t) & = \sum^\infty_{n=1} S_n(x) \chi_n^{S}(t) 
\end{aligned}
\vspace*{0.1cm}
\end{equation}
into the action \eqref{eq:action} and keeping quadratic terms in the fields. 

A possible way of determining the eigenmodes and eigenvalues of the mixed system is to diagonalize the 4D mass matrix resulting after integrating the quadratic terms in the action. Besides the KK scalar masses for the non-mixed case, non-diagonal entries arise through the potential, once we integrate the profiles over the fifth dimension, i.e.\
\begin{widetext}
\begin{equation}
\begin{aligned} 
\int d^5x \sqrt{g} \, V(H,\Sigma) \supset \bl \int d^4 x \int_\epsilon^1 \dfrac{dt}{t^3} \Bigg[ & \dfrac{\mkk^2}{kr} v_S^2(t) \, h(x,t)^2 + \dfrac{kr}{(\mu_S r)^2} \dfrac{\lambda_S}{r} t^2 \vp^2_H(t) S(x,t)^2 \\
& + 4 \dfrac{\mkk}{\mu_S r} \sqrt{\dfrac{\lambda_S}{r}} \, t \, \vp_H(t) v_S(t) h(x,t) S(x,t) \Bigg] .
\end{aligned} 
\end{equation} 
Inserting the KK decompositions for $h(x,t)$ and $S(x,t)$, and using the profiles for the decoupled case, we can write down the mass matrix to first order in $\bl$, finding
\begin{equation}
   \mathcal{M}^2 = \!
		\left[
\begin{pmatrix} \vspace*{0.1cm}
	x_{h_0}^{2} & 0 & 0 & \cdots \\ \vspace*{0.1cm}
	0 & x_{S_1}^{ 2} &0 & \cdots\\
	0 & 0 & x_{h_1}^{2} & \cdots \\
	\vdots & \vdots & \vdots & \ddots
\end{pmatrix} \! + \bl 
\begin{pmatrix} \vspace*{0.1cm}
	\kappa_{h_0}^{2} & \kappa^{2}_{h_0 S_1}& \kappa^{2}_{h_0 h_1} & \cdots\\ \vspace*{0.1cm}
	\kappa^{2}_{h_0 S_1} & \kappa_{S_1}^{ 2}&\kappa^{2}_{h_1 S_1} & \cdots\\ 
	\kappa^{2}_{h_0 h_1}&\kappa^{2}_{h_1 S_1} &\kappa^{ 2}_{h_1} & \cdots \\
	\vdots & \vdots & \vdots & \ddots
\end{pmatrix}\right] \mkk^2.
\label{eq:mass_matrix}
\end{equation}
\end{widetext}
Here, $x_{h_n}$ and $x_{S_n}$ correspond to the unperturbed mass eigenvalues in units of $\mkk$ of the decoupled system (for $\bar\lambda=0$), defined as $x_i^2=m_{i}^2/\mkk^2$. Note that only $x_{h_0}\ll 1$ corresponds to a light zero mode, because the odd scalar field has no zero modes. With the mixing switched on, the mass of the lightest scalar becomes 
\begin{equation}
	m_h^2 \approx \left( x_{h_0}^{ 2} + \bl \, \kappa_{h_0}^{2} \right) \mkk^2 \, . \label{eq:mass_higgs}
\end{equation}
In order to calculate the $\mathcal{O}(\bar\lambda)$ terms we need the unperturbed solutions for the profiles. For the Higgs KK modes they are given by
\begin{equation}
	\chi_n^h(t) = \sqrt{\dfrac{L}{\pi}} \dfrac{t J_\beta(x_{h_n} t)}{\sqrt{J^2_\beta(x_{h_n}) - J_{\beta+1}(x_{h_n}) J_{\beta-1}(x_{h_n})}} \, ,
\end{equation}
where $J_\beta(x)$ is a Bessel function and the eigenvalues $x_{h_n}$ satisfy 
\begin{equation}
	\dfrac{x_{h_n} J_{\beta + 1} (x_{h_n} )}{J_{\beta} (x_{h_n} )} = 2 \left( \mir -2 -\beta \right) \equiv 2 \delta. \label{eq:bc_higgs}
\end{equation}
At zeroth order in $\bl$ and for $x_{h_0}^2 \ll 1$ the expression for the zero-mode profile $\chi_0^h(t)$ is approximately given by
\begin{equation}
\chi_0^h(t) \simeq \sqrt{\dfrac{L}{\pi}(1+\beta)} \, t^{1+\beta},
\label{eq:higgs_profile}
\end{equation}
up to $\mathcal{O}(x_{h_0}^2)$ corrections.

The contributions to the mass matrix at first order in $\bl$ can be read from
\begin{equation}
\begin{aligned} 
		& \kappa_{S_m}^{2} = \dfrac{2\, kr}{(\mu_S r)^2} \dfrac{\lambda_S}{r} \int_\epsilon^1 \dfrac{dt}{t} \dfrac{\vp^2_H(t)}{\mkk^2} [\chi_m^S(t)]^2 \, ,\\
	& \kappa_{h_nS_m}^{2} = \dfrac{4}{\mu_S r} \sqrt{\dfrac{\lambda_S}{r}} \int_\epsilon^1 \dfrac{dt}{t^2} \dfrac{\vp_H(t)}{\mkk} v_S(t) \chi_n^h(t) \chi_m^S(t) \, , \\
& \kappa_{h_n h_m}^2 = \dfrac{2}{kr} \int_\epsilon^1 \dfrac{dt}{t^3} v_S^2(t) \chi_n^h(t)\chi_m^h(t) \, ,\\
\end{aligned} 
\end{equation} 
where $\kappa_{h_n}^2\equiv \kappa_{h_n h_n}^2$ in (\ref{eq:mass_matrix}). The different powers of $t$ in the denominator result from our particular normalization of the VEV of the new scalar field in (\ref{Sigmadecomp}), which differs from the normalization of the Higgs VEV in (\ref{eq:higgsfield}).

A priori, both $x_{h_0}$ and $\kappa_{h_0}$ are naturally ${\cal O}(1)$ numbers, so in order to obtain a 125 GeV Higgs boson one needs to tune
\begin{equation}\label{eq31}
	\frac{m_h^2}{\mkk^2} \approx x_{h_0}^{ 2} + \bl \, \kappa_{h_0}^{2} \sim \left(\frac{0.125}{5}\right)^2\sim 10^{-3}. 
\end{equation}
This is a well-known feature of bulk Higgs models in WEDs~\cite{Davoudiasl:2005uu, Cacciapaglia:2006mz, Azatov:2009na, Vecchi:2010em, Archer:2012qa, Malm:2013jia, Archer:2014jca}. 
We can achieve this in two different ways. For positive values of $\bar{\lambda}$, both terms in the sum need to be small simultaneously. In the case of $x_{h_0}^2$, this can be achieved by tuning the parameters in the Higgs potential, as it is customary for a bulk Higgs with no additional scalars~(see e.g. \cite{Malm:2013jia, Archer:2014jca}). For $\bl\, \kappa_{h_0}^2$ the only possibility is to make $\bl$ small enough, since $\kappa_{h_0}^2$ is an ${\cal O}(1)$ number unless very large values of $\beta$ are chosen. (The limit $\beta\to\infty$ corresponds to a brane-localized Higgs and will not be considered here.) Therefore, for positive values of~$\bl$ 
\begin{equation}
0\le	\bl \, \kappa_{h_0}^2 \sim \bl \lesssim 10^{-3}. 
\end{equation}
As a result, in this case values of $\bl$ larger than $10^{-3}$ are not allowed, regardless of the value for $x_{h_0}^2$.

\begin{figure*}[t!]
\begin{center}
\hspace*{-.25cm}
\includegraphics[scale = 0.45]{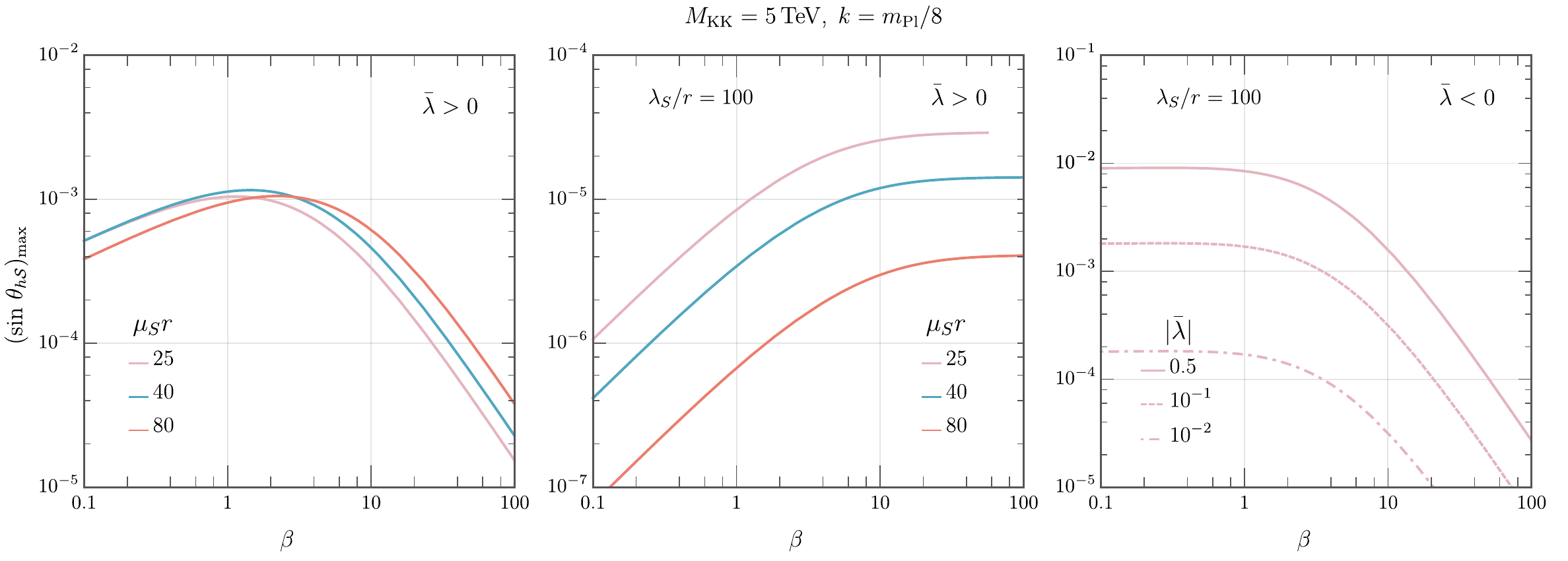}
\vspace*{-0.25cm}
	\caption{ Maximum allowed value for the parameter $(\sin\theta_{h\mathcal{S}})_{\max}$ describing the mixing between the lightest Higgs mode and the first KK resonance of the $\mathbb{Z}_2$-odd scalar as a function of $\beta$, for fixed values of $\mkk$ and $k/m_{\rm Pl}$. In the left and middle plots, we show this dependence for different values of $\mu_S r$ and positive $\bl$. In the left plot, we consider the maximum possible values of $\bl>0$ and $\lambda_S/r$, whereas we fix $\lambda_S/r$ to $100$ in the middle plot,  with $\bl$ still saturating the resulting upper bound. Finally, in the right plot we fix both $\mu_S r$ and $\lambda_S/r$ and consider three different negative values of $\bl$.} \label{fig:plotmixing}
\end{center}
\end{figure*}

One could also entertain the possibility of considering solutions in which both quantities $x_{h_0}^2$ and $\kappa_{h_0}^2$ are simultaneously $\mathcal{O}(1)$, but they cancel each other out leading to a light Higgs mass. Since $\kappa_{h_0}^2>0$ by definition, one would need to have either $x_{h_0}^2$ or $\bl$ negative. The first possibility corresponds to a tachyonic Higgs field (before turning on the mixing with the odd scalar) and leads to the BC 
\begin{equation}
	\dfrac{x_{h_n} I_{\beta + 1} (x_{h_n} )}{I_{\beta} (x_{h_n} )} = -2 \left( \mir -2 -\beta \right) \equiv -2 \delta
\end{equation}
on the IR brane, where $I_n(x)$ are modified Bessel functions. 
This condition is similar to that in \eqref{eq:bc_higgs}, but with a relative minus sign. However, such a path leads nowhere since, as can be proven, this equation is incompatible with the presence of a Higgs VEV ~\cite{Archer:2012qa, Archer:2014jca}. Therefore, the only viable option is to allow for negative values of $\bl$. In that case, equation~\eqref{eq:mass_higgs} becomes
\begin{equation}
	m_h^2 \approx \left(x_{h_0}^2 - |\bl|\, \kappa_{h_0}^2\right)\mkk^2 . \label{eq:mhneglamb}
\end{equation}
and we can always reproduce the Higgs mass regardless of the value of $\bl<0$, by choosing the appropriate value of $x_{h_0}^2$. 

At any rate, the mixing between the even and odd bulk scalars leads to
\begin{equation}
		h_{0}(x)=h_{\rm phys}(x)+\sin\theta_{h\mathcal{S}}\, \mathcal{S}(x)+\sin\theta_{h\mathcal{H}}\, \mathcal{H}(x),
\end{equation}
where $\mathcal{H}(x)=h_1(x)+\mathcal{O}(\bl)$ and $\mathcal{S}(x)=S_1(x)+\mathcal{O}(\bl)$ are the profiles of the first KK modes in the limit where $\bar\lambda=0$, and
\begin{equation}
	\begin{aligned}
			\sin \theta_{h\mathcal{S}} &= \bl \dfrac{\kappa_{h_0 S_1}^2}{x_{S_1}^2 - x_{h_0}^2} \approx \bl \dfrac{\kappa_{h_0 S_1}^2}{x_{S_1}^2} \, ,\\
		\sin \theta_{h\mathcal{H}} &= \bl \dfrac{\kappa_{h_0 h_1}^2}{x_{h_1}^2 - x_{h_0}^2} \approx \bl \dfrac{\kappa_{h_0 h_1}^2}{x_{h_1}^2} \, .
		\label{eq:senos}
	\end{aligned}
\end{equation}
In general, the mixing of the lightest Higgs eigenmode with the first odd excitation can be expressed as
\begin{equation} 
\sin \theta_{h\mathcal{S}} \simeq 4 \bl \sqrt{\dfrac{\lambda_S}{r}} \dfrac{x_4}{x_{S_1}^2} \dfrac{kr}{\mu_S r} (1+\beta) \int_\epsilon^1 dt \; t^{2\beta} v_S(t) \chi_1^S(t) \, ,
\end{equation}
and a similar expression can  be derived for $\sin \theta_{h\mathcal{H}}$, i.e., 
\begin{equation}
	\sin\theta_{h\mathcal{H}}\simeq \dfrac{2\bl}{ x_{h_1}^2}\sqrt{\dfrac{1+\beta}{kr}}  \int_\epsilon^1 dt\, t^{\beta-2} v_S^2(t) \chi_1^h(t).
\end{equation}

As we can see, when $\bl$ is positive the constraint set by the physical Higgs mass does not allow for a large mixing. Its upper bound is saturated when one assumes that the leading contribution to the Higgs mass is given by the $\bl\, \kappa_{h_0}^2$ term in (\ref{eq31}). Then 
\begin{equation}
	\bl_{\rm max} = \dfrac{x_h^2}{\kappa_{h_0}^2} = \dfrac{x_h^2}{2(\beta+1)} \left[ \int_\epsilon^1 dt \; t^{2\beta-1} v_S^2(t) \right]^{-1}.
	\label{eq:blmax}
\end{equation}
In this case, plugging in the expression for $\kappa_{h_0 S_1}^2$ and $x_{S_1}^2$, we get 
\begin{equation}\label{eq39}
 \hspace*{-0.25cm}
(\sin \theta_{h\mathcal{S}})_{\rm max} \simeq 2 \dfrac{x_h^2 x_4}{x_{S_1}^2} \dfrac{k}{\mu_S} \sqrt{\dfrac{\lambda_S}{r}} \dfrac{\int_\epsilon^1 dt \; t^{2\beta} v_S(t) \chi_1^S(t) }{\int_\epsilon^1 dt \; t^{2\beta-1} v_S^2(t) } \, ,
\end{equation}
where   $\lambda_S / r$ in the above equation needs to saturate its upper bound  \begin{equation}
\dfrac{\lambda_S}{r} \le \dfrac{\mu_S^4}{k^4} \dfrac{2}{kr \, x_4^2 \, x_h^2} \int_\epsilon^1 dt \; t^{2\beta-1} v_S^2(t) ,
\end{equation}
which results after inserting \eqref{eq:blmax} into equation~\eqref{eq:max_condition_2}.  This leads to 
\begin{equation} \hspace*{-0.25cm}
(\sin \theta_{h\mathcal{S}})_{\rm max} \simeq \dfrac{x_h}{x_{S_1}^2} \dfrac{2^{3/2} \, \mu_S r}{\left(kr\right)^{3/2}} \dfrac{\int_\epsilon^1 dt \; t^{2\beta} v_S(t) \chi_1^S(t) }{\left(\int_\epsilon^1 dt \; t^{2\beta-1} v_S^2(t) \right)^{1/2}} \, ,
\end{equation}
which only depends on $\beta$ and $\mu_S r$, given that $kr\sim\mathcal{O}(10)$ in order to solve the hierarchy problem and that the eigenvalues $x_i$ and the scalar profiles are determined once these parameters have been fixed.

When $\bl$ is negative, on the other hand, $\lambda_S/r$ is unconstrained by relation \eqref{eq:max_condition_2}. In this case, an upper bound on $\lambda_S/r$ arises if one wants to prevent the theory from becoming strongly coupled, since the couplings of the KK scalar field $\mathcal{S}$ to the different fermions are proportional to $\sqrt{\lambda_S/r}$, as one can see from equations~\eqref{eq:oddyuk} and \eqref{eq:oddyukdm} in the appendix. Moreover, in this case sizable values of $|\bl|\sim \mathcal{O}(1)$ are allowed. For all these reasons, we find that $\sin\theta_{h\mathcal{S}}$ can be much larger than in the case of a positive~$\bl$. 

In figure~\ref{fig:plotmixing}, we show the different predictions for the maximum allowed value of the parameter $(\sin \theta_{h\mathcal{S}})_{\max}$, which measures the mixing between the lightest Higgs scalar and the first KK mode of the new $\mathbb{Z}_2$-odd scalar as a function of $\beta$. In the left plot, we show this dependence for different values of $\mu_S r$ 
after saturating the upper bounds on $\lambda_S/r$ and $\bl>0$. In the middle plot, we display the maximum allowed value of $\sin\theta_{h\mathcal{S}}$ for the same values of $\mu_S r$ and a fixed value $\lambda_S/r=100$,  together with $\bl=\bl_{\max}>0$. Note that for $\mu_S r=25$ and $\beta\sim 50$, $\lambda_S/r=100$ takes its maximum value. This explains why, in this case, the line stops before one can reach $\beta=100$. Finally, in the right plot we show $(\sin\theta_{h\mathcal{S}})_{\max}$ for different values of $\bl<0$ 
and fixed values $\mu_S r=25$ and $\lambda_S/r=100$. In all these plots we assume $\mkk=5$~TeV and $k=m_{\rm Pl}/8$. One can readily see that, for a given value of $\lambda_S/r$, the maximum allowed value for $\sin\theta_{h\mathcal{S}}$ is much more significant in the case $\bl<0$, since larger values of $|\bl|$ can be taken. In addition, when $\bl$ is negative one could also consider bigger values of $\lambda_S/r$ than in the $\bl>0$ case.  All this results into larger mixing angles when $\bl$ is negative compared to the  $\bl>0$ case.

\begin{figure*}[t!]
\begin{center}
\hspace*{-.25cm}
\includegraphics[scale = 0.47]{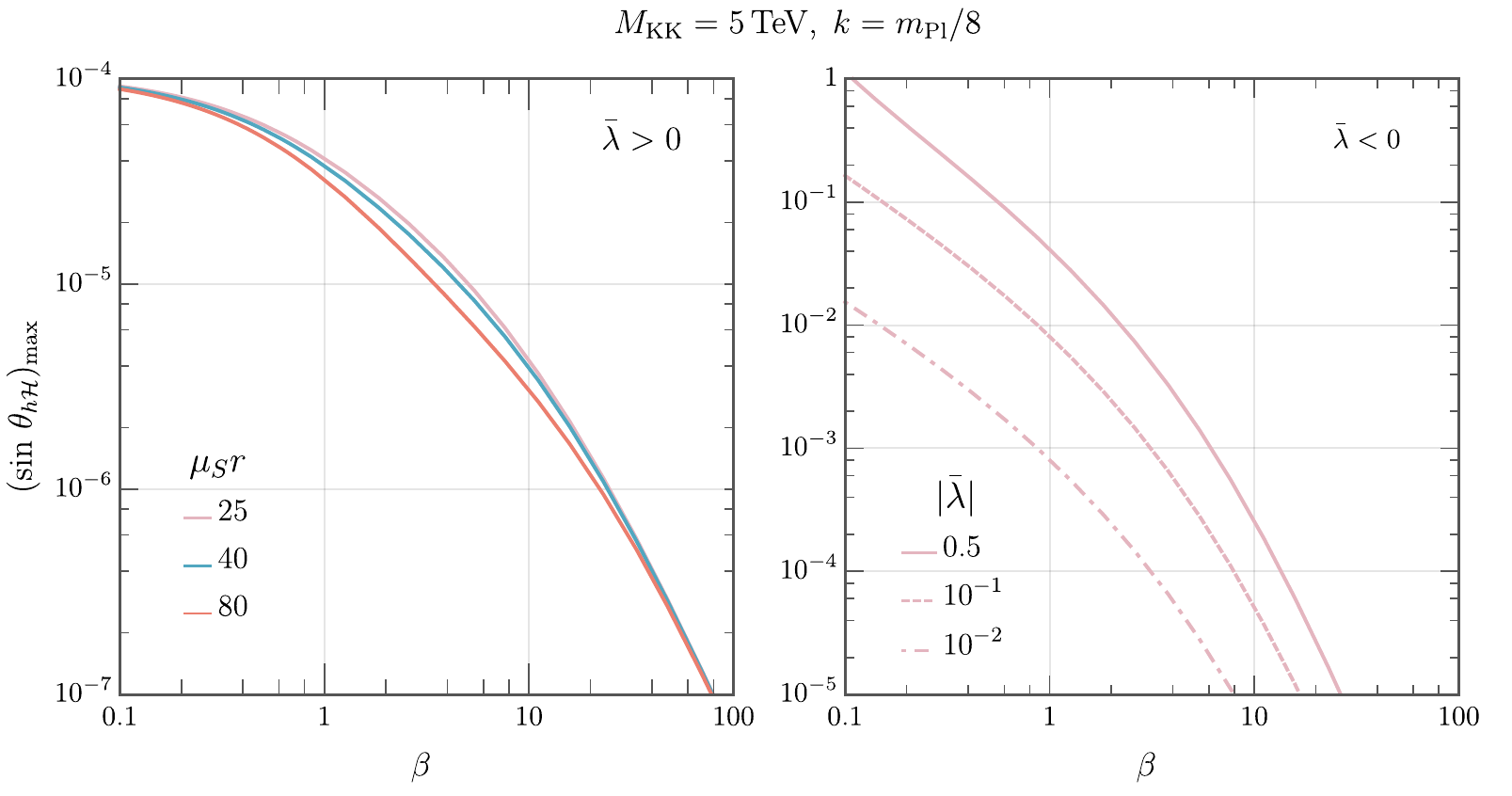}
\vspace*{-0.25cm}
\caption{Maximum allowed value for the parameter $\sin\theta_{h\mathcal{H}}$, describing the
	mixing between the lightest Higgs mode with its first KK resonance, as a function of $\beta$. In the left figure, we exhibit three different values of $\mu_S r$  for $\bl=\bl_{\max}>0$, while in the right panel we consider three different negative values of $\bl$ for a fixed value of $\mu_S r=25$. } \label{fig:plotmixingH}
\end{center}
\end{figure*}

For the Higgs mixing with its first KK mode, parametrized by $\sin\theta_{h\mathcal{H}}$, we found a monotonic behavior, with $\sin \theta_{h\mathcal{H}}$ getting smaller for large values of $\beta$ independently of the $\mu_Sr$ parameter. This can be seen in figure~\ref{fig:plotmixingH}, where we show $(\sin\theta_{h\mathcal{H}})_{\max}$ as a function of $\beta$ for $\mkk=5$~TeV and $k=m_{\rm Pl}/8$. In particular, we display on the left panel this functional dependence for three different values of $\mu_S r$, after saturating $\bl$ to its upper bound. 
In the right panel we exhibit the case where $\mu_S r=25$ is kept fixed, while $\lambda<0$ takes on different (negative) values. 
Comparing this panel with the right panel of the previous figure, we can see that $(\sin \theta_{h\mathcal{H}})_{\max}$ is a steeper function of $\beta$ than $(\sin\theta_{h\mathcal{S}})_{\max}$ when $\bl<0$. For the choice of parameters at hand, and depending on the value of $|\bl|$, $\sin \theta_{h\mathcal{H}}$ becomes bigger than $\sin\theta_{h\mathcal{S}}$ for $\beta$ smaller than values between $1$ and $10$, depending on the other parameters of the model, whereas the opposite happens  when $\beta$ takes larger values. Henceforth, for practical purposes, we will neglect the differences between $\mathcal{S}$ and $S_1$, as well as between $\mathcal{H}$ and $h_1$, since they are proportional to the small mixing angles $\sin\theta_{h\mathcal{S}}$ and $\sin\theta_{h\mathcal{H}}$, respectively.

\section{Phenomenology}\label{sec:pheno}

Once the Higgs boson is allowed to propagate into the bulk of the extra dimension, its mixing with the $\mathbb{Z}_2$-odd scalar becomes unavoidable. This mixing will leave its imprint on different aspects of the phenomenology. In particular, it can lead to effects on experiments as diverse as high-energy colliders, both present and future ones. Moreover, as we will see, assuming the presence of dark fermions, it can naturally explain the observed DM relic abundance and leave its imprint on DM direct-detection experiments.

We have seen how the quartic coupling leads to a mass mixing of the Higgs-boson zero mode with both its first KK resonance $h_1$ and the lowest-lying $\mathbb{Z}_2$-odd scalar, $S_1$. This mixing induces modifications on the Higgs-boson couplings to SM particles. In section~\ref{subsec:modf} we will explore these modifications and study its impact on current and future colliders.

A key aspect of our model is that the odd scalar field constitutes a unique window into dark sectors featuring fermions propagating into the bulk. Indeed, since all the 5D fermion bulk masses are generated through Yukawa-like interactions with the odd scalar, the scalar KK modes necessarily connect any dark fermionic sector with the SM if the former is genuinely five dimensional. Such a connection is unavoidable and constitutes a defining feature of the model. In the presence of a dark fermionic sector, the required Yukawa couplings between the odd scalar field and the bulk fermions has two interesting consequences. On the one hand, for light enough dark fermion masses, it induces a Higgs invisible decay width, and this in turn implies constraints on the size of the scalar mixing between both 5D scalar fields. We study this in detail in section~\ref{subsec:invisible}. On the other hand, the dynamical generation of the 5D fermion masses naturally connects the visible and the invisible sectors via the KK resonances of the odd scalar field, with $S_1$ giving the leading contribution. In particular, this introduces an efficient coannihiliation channel for the lightest dark fermion, which is naturally stable and therefore a good DM candidate. We study this possibility both in the regular freeze-out scenario and in the case of a matter-dominated freeze-out in section~\ref{sec:DM}. Finally, in section~\ref{sec:xenon} we study in detail the constraints coming from direct-detection experiments using recent Xenon1T data~\cite{DiGangi:2018xht, Aprile:2018dbl}.

	\subsection{Modified Higgs couplings}\label{subsec:modf}

As we have seen in the previous section, the physical Higgs boson can be expressed with very good approximation as a linear combination of the interaction eigenstates $h_0$, $h_1$ and $S_1$. Since these interaction eigenstates couple differently to the SM particles, this mixing induces modifications of the SM Higgs couplings. Here, we study the implications of these modifications.

The 4D effective couplings of the different scalars to the fermions are obtained by integrating the profiles of the different fields over the fifth dimension and a subsequently rotate into the mass basis. In particular, the coupling of the Higgs-boson zero mode and KK modes to a pair of fermion chiral zero modes, $\bar \Psi_a \Psi_b$, is given by
\begin{equation}
	y_{abh_n} = \dfrac{y_{\ast}}{ \sqrt{kr}}\dfrac{2+\beta}{\sqrt{2(1+\beta)}} \int_\epsilon^1 dt f_a (t) f_b(t) \chi_n^h(t) \,,
	\label{eq:yukhs}
\end{equation}
where $y_{\ast}$ is defined as a function of the 5D dimensionful Yukawa coupling $Y_{\rm 5D}$ \cite{Malm:2013jia} 
\begin{equation}
	y_{\ast}=\frac{\sqrt{k(1+\beta)}}{2+\beta}Y_{\rm 5D},
\end{equation}
where the latter is defined by (for an up-type quark field $\Psi_{Rb}$, the Higgs-boson field $H$ must be replaced by $\tilde H$)
\begin{equation} 
	 	S_Y\supset  -\int d^5 x\sqrt{g}\,Y_{\rm 5D}\bar\Psi_{La}(x,t)H(x,t)\Psi_{R b}(x,t) + \mathrm{h.c.} \,.
\end{equation}
On the other hand, the coupling of two light SM fermions to the scalar $\mathcal{S}$  only appears through a mass insertion. Indeed, before EWSB there is no direct coupling between $\mathcal{S}$ and two SM-like fermion fields. Such a coupling is only generated after taking into account the fermion mixing induced by the Higgs VEV $v_{\rm SM}$. We will compute the corresponding coupling perturbatively, as it is expected to be suppressed by a factor of $\mathcal{O}(v_{\rm SM}/\mkk)$. This coupling arises from the interactions of the $\mathcal{S}$ scalar to the different fermion zero modes and the first KK resonance with opposite chirality, once we rotate to the fermion mass basis after EWSB. Specifically, the coupling between $\mathcal{S}$, a chiral fermion zero mode $a$, and its first KK resonance $A$ with opposite chirality, is given by
\begin{equation}
	\begin{aligned}
		y_{aA\mathcal{S}}& = 2 \, c_a \sqrt{\dfrac{\lambda_S}{r}} \dfrac{k}{\mu_S} \int_\epsilon^1 dt f_a(t) f_A^R(t) \chi_1^S(t),\quad \mathrm{or}\\
		y_{Aa\mathcal{S}}& = 2 \, c_a \sqrt{\dfrac{\lambda_S}{r}} \dfrac{k}{\mu_S} \int_\epsilon^1 dt f_A^L(t) f_a(t) \chi_1^S(t),
	\label{eq:syuk}
	\end{aligned}
\end{equation}
 depending on the zero-mode chirality, where $c_a$ is defined in appendix~\ref{app:fermions}. Then, after rotating the fermion fields to the mass basis, we induce an interaction term $y_{f\mathcal{S}}\mathcal{S}\bar{f}_Lf_R$ between the  SM-like chiral fields, $f_L$ and $f_R$,  and $\mathcal{S}$. 

We can write 
\begin{equation}
\begin{aligned}
	\delta y^{\rm phys}_{fh} &\equiv 1-\frac{y^{\rm phys}_{fh}}{y_{fh}^{\rm SM}}\simeq (1-\varkappa_f) +\Delta_{f\mathcal{H}} + \Delta_{f\mathcal{S}},
\end{aligned}
\end{equation}
where we have defined $\varkappa_f=y_{f_L f_R h_0}/y_{fh}^{\rm SM}$. Here $y_{fh}^{\rm phys}$ is the resulting Higgs Yukawa coupling 
\begin{align}
	\mathcal{L}\supset -\frac{1}{\sqrt{2}}y_{fh}^{\rm phys} h \bar{f}_Lf_R+\mathrm{h.c.},
\end{align}
and $y_{fh}^{\rm SM}$ denotes the corresponding parameter in the SM one. Moreover 
\begin{equation}\label{eq48}
	\Delta_{f \mathcal{H}}=\sin\theta_{h\mathcal{H}} \dfrac{y_{f_L f_R h_1}}{y_{f_L f_R h_0}}, \quad 	\Delta_{f \mathcal{S}}=\sin\theta_{h\mathcal{S}} \dfrac{y_{f\mathcal{S}}}{y_{f_L f_R h_0}}.
\end{equation}
The quantity $\varkappa_f$ measures the ratio of the Yukawa coupling of the Higgs-boson zero mode relative to that of the SM Higgs boson, whereas the parameters $\Delta_{f \mathcal{H}}$ and $\Delta_{f\mathcal{S}}$ describe the admixtures of the Higgs-boson and $\mathbb{Z}_2$-odd scalar KK modes into the physical Higgs.
Note that we have taken $y_{fh}^{\rm SM}$ equal to $y_{f_L f_R h_0}$ in the denominator of $\Delta_{f\mathcal{H}}$ and $\Delta_{f\mathcal{S}}$, since the difference is $\mathcal{O}(\bl\, v_{\rm SM}^2/\mkk^2)$ and thus subleading. On the other hand, $\varkappa_f$ is blind at this order to the Higgs mixing with the odd scalar. This is a byproduct of the induced light-heavy fermion mixing after EWSB and the shift in the 5D Higgs VEV. It has been studied e.g. in \cite{Archer:2014jca} for the RS case. In particular, for $\mkk=5$~TeV, $\varkappa_b$ never exceeds $2.5\cdot 10^{-2}$ when $1\le \beta \le 10$. In the case of lighter quarks, even smaller values are expected. In this work we concentrate on $\Delta_{f\mathcal{S}}$ and $\Delta_{f\mathcal{H}}$, since they are direct probes of the mixing of the bulk Higgs field with the $\mathbb{Z}_2$-odd scalar field.

\begin{figure}[t!]
\begin{center}
\hspace*{-.25cm}
\includegraphics[scale = 0.5]{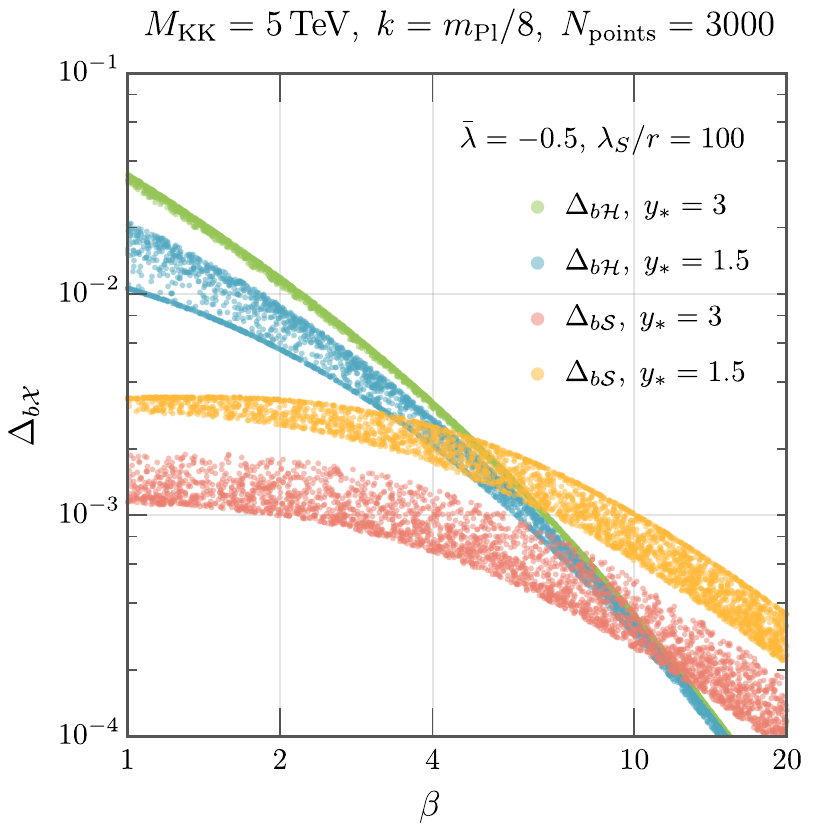}
\vspace*{-0.25cm}
\caption{$\Delta_{b\mathcal{H}}$ and $\Delta_{b\mathcal{S}}$ as functions of $\beta$ for two different values of $y_{\ast}$ and fixed values of $\bl$, $\lambda_S/r$, $\mkk$ and $k/m_{\rm Pl}$. For each case, we have generated $N_{\rm points}=3000$ random values of $c_{t_R}\in [-0.6, 0.2]$ and obtained $c_{q_L^3}$ and $c_{b_R}$ by correctly reproducing the top- and bottom-quark masses.} \label{fig:plotmody}
\end{center}
\end{figure}

Hereafter, we will focus on the bottom quark. The reasons for this are twofold. On the one hand, we expect the modifications of the Yukawa couplings to be larger for heavier quarks, while on the other hand, the bottom Yukawa coupling is among those measured most accurately, having in addition the most promising prospects. Indeed, existing measurements of the $h\to b\bar{b}$ signal strength (relative to the SM expectation) lead to $\mu_{h\to bb}=1.01 \pm 0.12\, (\mathrm{stat}.)\,^{+0.16}_{- 0.15}\, (\mathrm{syst}.)$~\cite{201859}. Assuming SM production this translates into a measurement of $(y_{bh}^{\rm phys}/y_{bh}^{\rm SM})^2$ with an uncertainty of about  $20\%$. However, the expected relative precision to be reached at future particle colliders such as the ILC, CLIC and the FCC is projected to be about 1\% for the initial stages, reaching the 0.5\% level in later stages of the experiments~\cite{deBlas:2019rxi}.

Figure~\ref{fig:plotmody} shows a scatter plot with values of $\Delta_{b\mathcal{H}}$ and $\Delta_{b\mathcal{S}}$ as a function of $\beta$, for $\bl=-0.5$ and $\lambda_S/r=100$. For both quantities we display two different scenarios, corresponding to $y_{\ast}=3$ and $y_{\ast}=1.5$, where $y_{\ast}$ is taken the same for both third-generation quarks (i.e., $y_{\ast}=y_{\ast}^t=y_{\ast}^b$). For each case, we have generated random values of $c_{t_R}\in[-0.6,0.2]$, and obtained $c_{q_L^3}$ and $c_{b_R}$ by fitting the top- and bottom-quark masses. One can see that for small values of $\beta$ the impact on $\Delta_{b\mathcal{H}}$ of changing $y_{\ast}$ is magnified. This is expected since, for these values of $\beta$, the Higgs is less IR-localized and  one  expects larger differences between the profiles $\chi_0^h$ and $\chi_1^h$. Indeed, regardless of the value of $y_{\ast}$, the prediction for the bottom mass $m_b\approx v_4\, y_{b_L b_R h_0}/\sqrt{2}$  needs to remain unchanged. Since $y_{b_L b_R h_0}$ is proportional to $y_{\ast}$, see~\eqref{eq:yukhs}, changes in $y_{\ast}$ have to be compensated by  different fermion localizations and therefore a different value of the overlap integral in that relation, in such a way that $y_{b_L b_R h_0}$ remains approximately constant.  The integral present in $y_{b_L b_R h_1}$ will change accordingly, but will deviate from the other integral as $\beta$ decreases.  This effect is reversed for $\Delta_{b\mathcal{S}}$, since $y_{\ast}$ is not explicitly present in the numerator of (\ref{eq48}), with the sole effect of changing the localization of the profiles of the third-generation fermions. Since a smaller value of $y_{\ast}$ requires a more IR-localized $q_L^3$ to reproduce the top-quark mass, one expects  a bigger overlap between $\mathcal{S}$ and the third-generation left-handed doublet $q_{L}^3$ together with its first KK mode. This leads to a larger value of $y_{b\mathcal{S}}$ and $\Delta_{b\mathcal{S}}$, as one can see in the figure. Note that $\Delta_{b\mathcal{S}}$ scales with  $\sqrt{\lambda_S/r}$, so one can readily obtain $\Delta_{b\mathcal{S}}$ for alternative choices of this parameter. Moreover, to good approximation both quantities scale linearly with $\bl$ for small values of $\bl$, because $\sin\theta_{h\mathcal{X}}\propto \bl$, as can be seen from \eqref{eq:senos}.

Taking into account the projected sensitivity for the bottom Yukawa coupling, one can see that we will be able to probe sizable values of the scalar portal coupling $\bl$ for moderately small values of $\beta$. In particular, as can be seen from figure~\ref{fig:plotmody}, for our chosen value $\bar\lambda=-0.5$, the predicted modifications $\Delta_{b\mathcal{H}}$ and $\Delta_{b\mathcal{S}}$ can be probed as long as $\beta$ is less than about 4. For smaller values of $\bl$, one could only access smaller values of $\beta$. 

\begin{figure}[t!]
\begin{center}
\hspace*{-0.1cm}
\includegraphics[scale = 1]{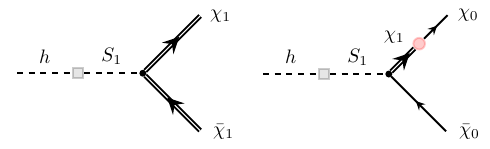}
\vspace*{-0.25cm}
\caption{Diagrams responsible for the generation of  the Higgs coupling to DM in the two main scenarios discussed in the text. The left figure corresponds to the case where the DM candidate is the first KK mode of a single 5D fermion field. The right figure corresponds to the case where the different chiralities of the DM candidate are zero modes of different 5D fields and the mediation via heavy KK modes is required. } \label{fig:diagramsdecay}
\end{center}
\end{figure}

The couplings of the Higgs KK modes to the $W$ and $Z$ bosons of the SM (corresponding to the zero modes of the corresponding bulk gauge fields) are proportional to the integrals
\begin{equation} 
g_{h_n\!V\!V} \propto \int_\epsilon^1 \!\dfrac{dt}{t}\varphi_H(t) \chi_n^h(t) \, ,
\end{equation}
where we have used that to a good approximation the zero-mode profiles for of the $W$ and $Z$ states are flat along the extra dimension \cite{Davoudiasl:1999tf,Pomarol:1999ad,Casagrande:2008hr}. Since $\varphi_H(t) \simeq v_4 \chi_0^h(t)$, and the orthogonality condition for the scalar profiles is given by \cite{Malm:2013jia, Archer:2014jca}
\begin{equation} 
\dfrac{2\pi}{L}\int_\epsilon^1 \dfrac{dt}{t} \chi_m^h(t) \chi_n^h(t) = \delta_{mn} ,
\end{equation}
the gauge bosons only couple to the Higgs-boson zero mode, and higher KK modes of the Higgs do not couple to the gauge-boson zero modes at tree level. For the case of the $S_1$ scalar, which is a SM singlet, the couplings to electroweak gauge bosons would only appear at the loop level. Therefore, it will not modify  the Higgs couplings to gauge bosons in an noticeable way, since these changes are suppressed by a small mixing angle   $\mathcal{O}(10^{-2})$ and a loop factor.

	\subsection{Invisible Higgs decays}\label{subsec:invisible}

\begin{figure*}[t!]
\begin{center}
\hspace*{-0.2cm}
\includegraphics[scale = 0.6]{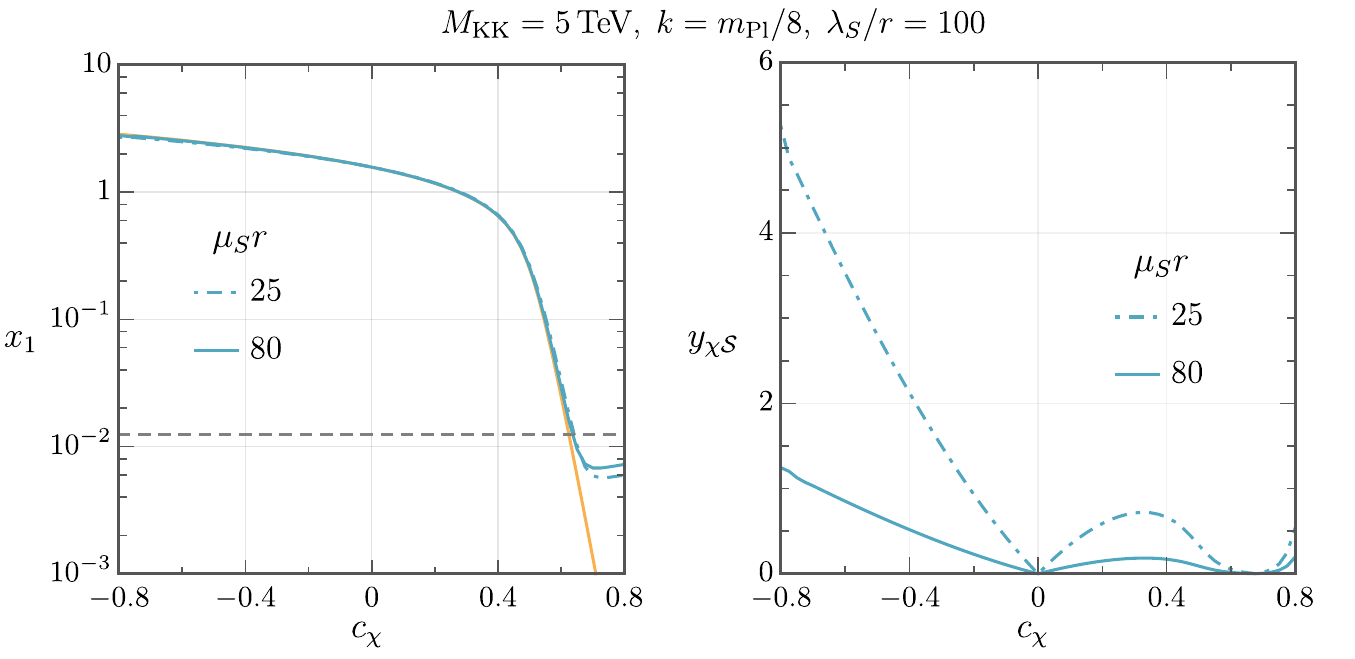}
\vspace*{-0.25cm}
	\caption{Left panel: Mass fraction $x_1=m_1/\mkk$ of the first fermion KK mode in terms of the   5D dimensionless bulk-mass parameter $c_{\chi}$ for the case of a left-handed chirality, with Dirichlet and Neumann boundary conditions on the UV and the IR brane, respectively, for two different values of $\mu_S r$. The yellow line corresponds to the case where the fermion bulk masses do not have a dynamical origin, but appear in the 5D Lagrangian along with a sign function, as it is usual in RS models.  We also show the value for which $m_\chi = m_h/2$ with a dashed gray line. Higgs decays into a pair of DM particles $\chi_1\bar\chi_1$ are kinematically allowed only if $x_1$ falls below this line. Right panel: Value of $y_{\chi\mathcal{S}}$ as a function of $c_{\chi}$ for the same choice of boundary conditions and values of $\mu_S r$. } \label{fig:plotcmasses}
\end{center}
\end{figure*}

The mixing between the Higgs and the $\mathbb{Z}_2$-odd bulk scalar field induces an effective coupling of the Higgs boson to any 5D bulk fermion present in the theory which is not localized on the UV or IR brane. This includes the possibility of fermions not charged under the SM group, the so-called dark fermions. Hereafter, we will consider this case and  will study its potential signatures, including its role in explaining the observed DM relic abundance. We will consider two different scenarios, depending on the origin of the dark fermion masses. In the first scenario, the dark fermion mass arises purely from orbifolding, i.e., from the compactification of the WED, and it is thus proportional to its curvature, $m_{\chi}\propto \mkk$. The simplest possibility is to add a 5D fermion with no zero mode, whose first KK resonance is automatically stable and a good DM candidate. In the case where one of the chiralities of the corresponding 5D field has a Dirichlet (Neumann) boundary condition on the UV (IR) brane, the first KK mode can be parametrically lighter than $\mkk$ \cite{Agashe:2004ci}, potentially light enough to be accessible in Higgs-boson decay. In this case, the DM candidate is VL even though it can have non-trivial quantum numbers under a possible new dark group. We thus allow for $N_{\chi}$ copies of such VL fermion. In the second scenario the fermion mass term connects two zero modes with opposite chiralities, arising from two different 5D fields. In this scenario, the fermion mass can either come through a dark Higgs mechanism or a brane-localized VL mass. If the dark Higgs is localized towards the IR brane, or alternatively the VL mass is localized on the IR brane, the fermion mass is proportional to $\mkk$, even though a large hierarchy can arise for UV-localized dark fermions. Either way, in this case $S_1$ does not interact directly with these two chiral zero modes  (in the same way $S_1$ does not couple directly to $b_L$ and $b_R$, as discussed  previously), since they are zero modes of two different 5D fields. Such a coupling can, however, be generated after a mass insertion through the mediation of the heavy KK modes (whether the mass comes from the spontaneous breaking of the dark gauge group or a VL mass is irrelevant to this discussion). We also assume a possible multiplicity of dark fermions given by $N_{\chi}$ as before.

We illustrate these two scenarios in figure~\ref{fig:diagramsdecay}. Note that in the second case we expect the coupling of the Higgs boson to the dark fermion to be $\mathcal{O}(m_{\chi}/\mkk)$ suppressed. Moreover, its calculation is rather model dependent. Alternatively, one could leave the nature of the dark fermion mass unspecified and define an effective Yukawa coupling, taking into account  the mixing of the heavy modes with the zero mode. 

There is an additional instance, which can be thought of as an intermediate scenario between the previous two cases. There one adds a 5D gauge-singlet fermion field with no additional flavor quantum numbers, which has a chiral zero mode and a Majorana mass term localized on the IR brane. The coupling of this field to $\mathcal{S}$ is generated analogously to the second case described above, via the involvement of a heavy KK mode and a mass insertion. Therefore, at the end of the day, this case is rather similar to the previous one, besides the difference in the multiplicities of fermionic degrees of freedom for a Majorana field. 

We have computed the mass of the first KK mode for a 5D field with mixed boundary conditions, as described in the first case above. This result is well known for non-dynamical bulk masses but has never been explored when these are generated by the VEV of a $\mathbb{Z}_2$-odd scalar. We show in the left panel of figure~\ref{fig:plotcmasses} the ratio $x_1=m_{1}/\mkk$  as a function of the usual dimensionless bulk-mass parameter $c$ in both scenarios. In the model at hand, where the different fermion bulk masses are generated by the VEV of the odd scalar field, $\langle \Sigma \rangle=\varphi_S,$  this parameter takes the form given in equation \eqref{eq:cobrakai}.  We also consider for comparison, the non-dynamical case where such bulk fermion masses are introduced by hand and read $c=m/k$, with $m$  the 5D bulk mass. In this scenario  the first KK mass $m_1$ can be made arbitrarily light by adjusting the $c$ parameter~\cite{Agashe:2004ci}. However, in our model a lower bound on the mass value arises due to behavior of $\varphi_S$ close to the two branes, where it vanishes. We find this bound to be $x_1 \sim 6 \cdot 10^{-3}$, corresponding to 30 GeV for our reference value $\mkk = 5$ TeV. We also show in the right panel of figure~\ref{fig:plotcmasses} the coupling of the VL fermions to the scalar $\mathcal{S}$, defined analogously to (\ref{eq:syuk}) but involving two KK profiles instead of one,
\begin{equation}
	y_{\chi\mathcal{S}} = 2 \, c_{\chi} \sqrt{\dfrac{\lambda_S}{r}} \dfrac{k}{\mu_S} \int_\epsilon^1 dt f_{1,\chi}^{L}(t) f_{1,\chi}^{R}(t) \chi_1^S(t),
	\label{eq:syukd}
\end{equation}
 as a function of $c_{\chi}$ for different values of $\mu_S r$. In both panels we have set $\mkk=5$ TeV, $k=m_{\rm Pl}/8$ and $\lambda_S/r=100$. We can see that, for large values of $m_{\chi}$, sizable values of $y_{\chi \mathcal{S}}$ are expected. Note that the VL fermions could also have a contribution to their mass coming from the dark sector (for instance an IR-localized Majorana mass term), however this would not affect their coupling to the scalar $\mathcal{S}$.  

The decay width of the Higgs boson into dark fermions is given by
\begin{equation}
	\Gamma(h \to \bar \chi \chi) = \dfrac{y_{\chi h}^2 N_\chi}{8\pi} m_h 
	\left( 1 - \dfrac{4 m_\chi^2}{m_h^2} \right)^{3/2} ,
\end{equation}
where we have defined 
\begin{equation}
	y_{\chi h} \equiv y_{\chi \mathcal{S}} \sin \theta_{h\mathcal{S}},
	\label{eq:hcoupdm}
\end{equation}
the coupling of the physical Higgs boson to the first dark KK fermion.

Using that $\mathcal{B}(H\to\text{inv}) < 0.33$ at 95\% CL \cite{Sirunyan:2018owy} and $\Gamma_H^{\text{SM}} \approx 4$ MeV, we can set an upper limit on the effective coupling of the Higgs boson to dark fermions. We find the upper bound on $y_{\chi h}$ to be $y_{\chi h} \lesssim 0.02/\sqrt{N_{\chi}} $ for DM candidates with mass $m_\chi < m_h/2$. Note that this constraint does not apply to heavier fermions.

	\subsection{Scalar-mediated fermionic dark matter} \label{sec:DM}

As discussed in the previous section, the $\mathbb{Z}_2$-odd scalar field will couple to any fermion field propagating in the bulk of the WED. This provides a robust bridge between the SM and any dark sector having fermions arising from 5D bulk fermion fields. In the case where these dark fermions are stable and make for a viable DM candidate, the KK excitations of the odd scalar field thus constitute efficient mediators for DM coannihilation into SM particles. Moreover, as we have already seen, the mixing between both scalar bulk fields induces a Higgs coupling to dark fermions, thereby turning the Higgs boson into an additional scalar mediator. 

\begin{figure}[t!]
\begin{center}
\hspace*{-0.7cm}
\includegraphics[scale = 1.1]{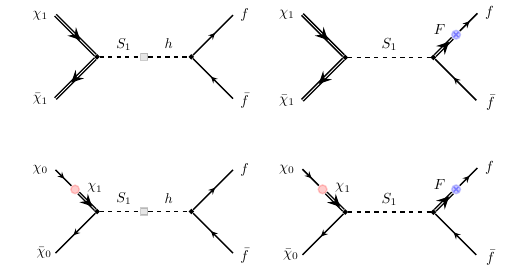}
\vspace*{-0.5cm}
	\caption{Diagrams contributing to the DM coannihilation cross section with fermions in the final state. The diagrams shown in the first line correspond to the case   where the DM candidate is a (potentially light) KK fermion $\chi_1$, whereas the diagrams in the second line correspond to the case where a mass insertion on a dark fermion line is needed in order to generate an effective interaction $S_1\bar{\chi}_0\chi_0$.} \label{fig:diagramsxsec}
\end{center}
\end{figure}

For the sake of concreteness, we will focus on the first scenario discussed in the previous section, i.e., of $N_{\chi}$ copies of a 5D dark fermion field, having potentially parametrically light KK modes. These potentially light KK modes -- the lightest dark particles -- are stable and a viable DM candidate. In this case, both the mass of the DM candidate $m_\chi$ and its couplings to the physical Higgs and the $\mathbb{Z}_2$-odd scalar, $y_{\chi h}$ and $y_{\chi \mathcal{S}}$, respectively, depend only on a single $c$ parameter (in addition to other model parameters such as e.g. $\mkk$, $k r$, $\lambda_S/r$ or $\mu_S r$), see figure~\ref{fig:plotcmasses}. Considering the alternative scenario where the interaction of $\mathcal{S}$ to the dark fermions requires a mass insertion on a dark fermion line, it would just lead to a different shape of the curve $y_{\chi S}=F(m_{\chi})$ and, by virtue of  \eqref{eq:hcoupdm}, also the value of $y_{\chi h}=\sin\theta_{h\mathcal{S}}\,y_{\chi S}$ (modulo a different count of degrees of freedom, in the case of Majorana fermions).  We illustrate in figure~\ref{fig:diagramsxsec} the diagrams relevant for the coannihilation $\bar{\chi}\chi \to \bar{f}f$ in these two cases, with and without a ``dark mass insertion'', as discussed in the previous section. We represent by a blue blob the heavy-light mass mixing induced after EWSB  in the visible sector, whereas the possible light-heavy mass mixing in the dark sector is depicted by a pink blob. Hereafter, for the sake of concreteness, we focus on the case of parametrically light KK fermions, for which there is no need of specifying any further dynamics in the dark sector.  

For Higgs-mediated processes, the dominant coannihilation final state will be $t\bar{t}$, if kinematically accessible (i.e.\ for $m_{\chi}>m_t$), or $b\bar{b}$, together with the vector final states $W^+ W^-$ and $ZZ$. In the case of diagrams mediated by $S_1$, $t\bar{t}$ or $b\bar{b}$ are the dominant coannihilation channels for moderately small values of $m_{\chi}$. However, for larger values of $m_{\chi}$, coannihilation into a SM fermion and its first KK resonance is also possible and can be the dominant coannihilation channel by far.

The relic abundance for a radiation-dominated freeze-out regime can be computed using~\cite{Gondolo:1990dk} (see also e.g.~\cite{Kim:2006af, Arcadi:2019lka})
\begin{equation}\label{eq54}
	\Omega_\chi h^2 \simeq \dfrac{x_f}{2\sqrt{g_{\star S}(m_\chi/x_f)}} \dfrac{10^{-9} \text{GeV}^{-2}}{\langle \sigma v \rangle},
\end{equation}
where $\Omega_\chi h^2 = 0.120 \pm 0.001$ \cite{Aghanim:2018eyx}. Here, $g_{\star S}(T_f)$ denotes the effective number of degrees of freedom in entropy  as function of the freeze-out temperature $T_f$, and we have defined a parameter $x_f=m_\chi/T_f$ to be determined below. $\langle \sigma v \rangle$ is the velocity-averaged cross section at the freeze-out temperature, which can be calculated as~\cite{Gondolo:1990dk} 
\begin{equation}
\begin{aligned}
\langle \sigma v \rangle & = \dfrac{1}{8 m_\chi^4 T_f K_2^2(m_\chi/T_f)} \times \\
	& \qquad\quad \int_{4m_\chi^2}^\infty\!ds\,\sigma(s) \left( s - 4 m_\chi^2 \right) 
	\sqrt{s} K_1(\sqrt{s}/T_f) ,
\end{aligned}
\end{equation}
where $K_n(x)$ are modified Bessel functions. The parameter $x_f$ in (\ref{eq54}) is obtained by solving the implicit equation
\begin{equation}
	x_f=\ln \left(g_{\chi}\frac{m_{\chi}}{2\pi^3}\sqrt{\frac{45}{8x_f\,g_{\star S}(m_\chi/x_f)}}\,
	m_{\rm Pl} \langle \sigma v\rangle \right) ,
\end{equation}
where $g_{\chi}=4 N_{\chi}$ is the number of DM degrees of freedom. 

\begin{figure*}[t!]
\begin{center}
\hspace*{-.25cm}
\includegraphics[scale = 0.55]{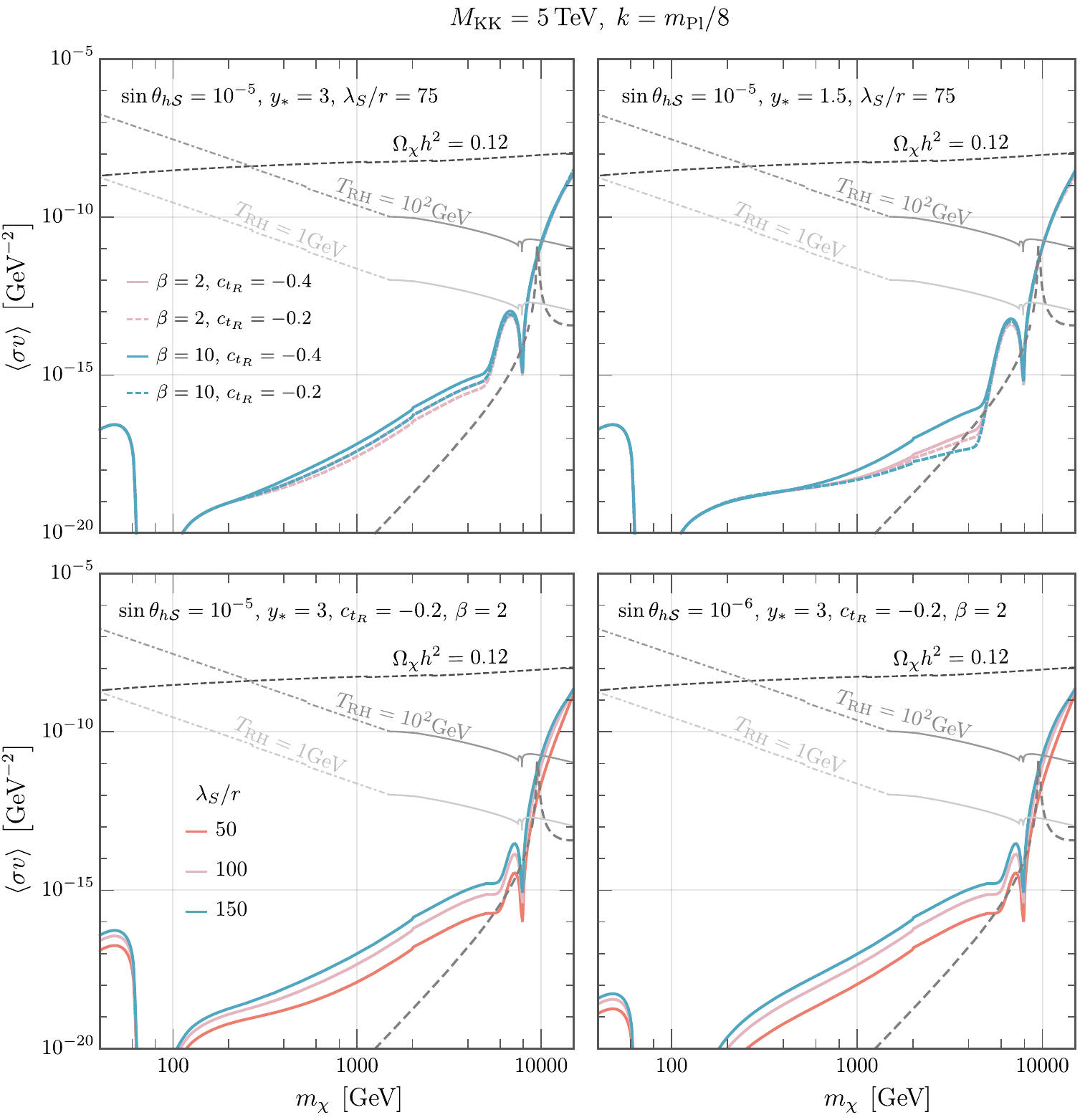}
\vspace*{-0.25cm}
	\caption{Velocity-averaged annihilation cross section $\langle \sigma v\rangle $ at the freeze-out temperature as a function of the DM mass $m_{\chi}$, for $N_{\chi}=1$, $\mkk=5$~TeV and $k=m_{\rm Pl}/8$. In the top panels, we fix  $\sin\theta_{h\mathcal{S}}$ and $\lambda_S/r$ and consider two different values of $y_{\ast}$. In both cases, we take two different values of  $\beta$ and $c_{t_R}$.  In the bottom panels, we fix $\beta$, $y_{\ast}$ and $c_{t_R}$ and consider different values of $\lambda_S/r$ for $\sin \theta_{h\mathcal{S}} = 10^{-5}$ (left) and $\sin \theta_{h\mathcal{S}} = 10^{-6}$ (right). In all four panels, we also show in dashed gray the $\langle \sigma v\rangle$ prediction for diagrams mediated by the exchange of the first KK graviton. We also show the velocity averaged cross section reproducing the relic density experimental value from Planck in dashed black, and the equivalent for a matter dominated freeze out in gray, for two different values of $T_{\rm RH}$, after using  $T_\star = 10^5$ GeV and $\tau=0.99$. For these lines the section in dot-dashed gray corresponds to predictions for which $x_f < 3$, and therefore in this regime the DM decouples relativistically \cite{Hamdan:2017psw,Chanda:2019xyl}.} \label{fig:cross_products}
\end{center}
\end{figure*}

Alternatively, one can also consider that DM freeze-out happens in an early period of matter domination, as proposed in~\cite{Hamdan:2017psw,Chanda:2019xyl}. Indeed, nothing prevents this from happening if radiation becomes dominant again before big-bang nucleosynthesis. The fact that DM decoupling happens during matter domination changes the freeze-out dynamics, since the Hubble rate has a different parametric dependence compared to the usual case, $H\propto T^{3/2}$ versus $H\propto T^2$. We do not elaborate here in detail on the dynamics behind this scenario, which is not crucial for our current analysis. One possibility would be to have a scalar field $\phi$ localized on the UV brane, which starts behaving like matter at a critical temperature $T_{\star}\sim m_{\phi}$ that we assume to be much larger than $\mkk$. If $\phi$ is sufficiently long-lived, its contribution to the energy density grows until it ultimately dominates the total energy density regardless of its initial contribution $(1-\tau)$ at $T_{\star}$, where $\tau\in[0,1]$ denotes the fraction of energy in radiation at $T=T_{\star}$. Following \cite{Hamdan:2017psw, Chanda:2019xyl} we will take $\tau=0.99$ as a benchmark value. Freeze-out happens at a temperature $T_f$, in a matter-dominated universe, before $\phi$ instantaneously decays at $T_{\rm \Gamma}<T_f<T_{\star}$, reheating the bath to $T_{\rm RH}$ and further diluting the DM freeze-out abundance. Hereafter we will assume $T_{\rm RH}\sim 1$ GeV. We refer the reader to appendix~\ref{app_mdf} for more details.

We show in figure~\ref{fig:cross_products} the velocity averaged coannihilation cross section $\langle \sigma v \rangle$ at the freeze-out temperature as a function of $m_\chi$, for $N_{\chi}=1$, $\mkk=5$~TeV and $k=m_{\rm Pl}/8$. In the top panels, we consider benchmarks with different values of $\beta$, $y_{\ast}$ as well as $c_{t_R}$ (the parameter fixing the localization of the RH top). In both top panels, we consider $\sin\theta_{h\mathcal{S}}=10^{-5}$ and $\lambda_S/r=75$, as well as two different values of $\beta$ and $c_{t_R}$. In particular, we show $\beta=2$ (pink), $\beta=10$ (blue), $c_{t_R}=-0.2$ (dashed line) and $c_{t_R}=-0.4$ (solid line). In the top-left panel we fix $y_{\ast}=3$ for both the up and the down third-generation quark sector, with $c_{q_L^3}$ and $c_{b_R}$ being determined by reproducing the top and bottom quark masses for a given choice of $c_{t_R}$. The same is done in the top-right panel but for $y_{\ast}=1.5$. In both cases, for the sake of simplicity, light quark masses are reproduced with UV localized fermions with identical bulk mass parameters (modulo a sign difference between opposite chiralities) and different values of $y_{\ast}$ with $y_{\ast}^s=1/2$ (for our purposes, such a not-so-refined study is more than enough). We can see that increasing the IR localization of the RH top, i.e. having bigger values of $|c_{t_R}|$, leads to a bigger cross section for most DM masses when $y_{\ast}=3$. Since $y_{\ast}$ is large enough in this case, changes in $c_{t_R}$ does not have a dramatic impact on $c_{q_L^3}$ and $c_{b_R}$, which remain almost unchanged. Therefore, the increase of the coannihilation cross section is mostly due to a larger $\mathcal{S}t_L t_R$ coupling, which is indeed the leading one for DM masses below about $10$~TeV. Such a larger coupling is the consequence of a bigger overlap with $\mathcal{S}$ and the increase in the Yukawa coupling $\mathcal{Y}$ coming with $c$. In the case of $y_{\ast}=1.5$, on the contrary, changes in $|c_{t_R}|$ do have a dramatic impact on $c_{q_L^3}$, since the RH top can not account for the top mass alone, requiring a fairly IR-localized third-generation quark doublet. Therefore, the contribution to $\mathcal{S} t_L t_R$ coming from the mixing of both top chiralites are similar, which leads to  bigger changes in the cross section in the region of DM masses between $1$ and $4$~TeV as one can see from figure~\ref{fig:cross_products} top-right panel. On the other hand, bigger values of $\beta$ lead in general to a larger mixing between fermion-zero modes and their KK resonances after EWSB, increasing the effective coupling $y_{f\mathcal{S}}$ after diagonalization. Therefore, in general, one expects a larger coannihilation cross section for $m_{\chi}\lesssim 10$~TeV and increasing values of $\beta$. Changing $\beta$ also affects the $\mathcal{S}t_L t_R$ coupling indirectly, since reproducing the observed quark masses results in different values of the mass parameters $c$. This explains why the dashed blue line in the top-right panel of figure~\ref{fig:cross_products} is below the other ones, since $c_{q_L}^3$ accidentally gets close to zero and thus reduces the left-handed doublet contribution to the $\mathcal{S}t_L t_R$ coupling, as can be seen in equation~\eqref{eq:syuk}.  Finally, note that the abrupt deep around $m_{\chi} \sim 8$~TeV is due to the zero in $y_{\chi \mathcal{S}}$ shown in figure~\ref{fig:plotcmasses}. Indeed, the cross section should exactly vanish at this point, but our numerical scan is unable to capture such an steep behavior.

In the bottom panels of figure~\ref{fig:cross_products}, on the other hand, we show $\langle \sigma v\rangle$ for different values of $\lambda_S/r$ as a function of $m_\chi$. In both bottom panels, we fix $\beta=2$, $y_{\ast}=3$ for both third-generation quark sectors, as well as $c_{t_R}=-0.2$. The left-bottom panel corresponds to the choice $\sin\theta_{h\mathcal{S}}=10^{-5}$, whereas for the bottom-right one we take $\sin\theta_{h\mathcal{S}}=10^{-6}$. By reducing the mixing, one effectively suppress the Higgs mediated contribution to the coannihilation cross section, which is mostly relevant for small DM masses and, in particular, around $m_{\chi}\approx m_{h}/2$. This will have an impact on direct detection as we will see later, since the Higgs provides the leading contribution to such experiments, and larger values of $\sin\theta_{h\mathcal{S}}$ will typically lead to more severe bounds from these experiments. The parameter $\lambda_S/r$ controls the effective Yukawa coupling of the $\mathcal{S}$ scalar to fermions $y_{\chi \mathcal{S}}$, see equations~\eqref{eq:syuk} and \eqref{eq:syukd}. We consider $\lambda_S/r=50$ (red), $\lambda_S/r=100$ (pink) and $\lambda_S/r=150$ (blue). Increasing $\lambda_S/r$ has the effect of increasing the coannihilation cross section in general, besides for values of $m_{\chi} \lesssim m_{\mathcal{S}}/2$ where the rise in the coupling is offset by the increase of its decay width. One should note that the resonant-like peak starting around $7-8$ TeV is not only due to the $\mathcal{S}$ resonance but also to the fact that new heavy-light final states become kinematically accessible in the coannihilation. They consist of a first KK fermion resonance of mass $\sim 15$ TeV together with a SM-like fermion. We do not show values of $m_{\chi}$ beyond $\sim 15$ TeV since the DM mass can not be made heavier than this value for $\mkk=5$ TeV. One could entertain the possibility of adding brane-localized masses or kinetic terms for this to happen, but for the sake of concreteness we do not explore such possibilities here. At any rate, for such large values of $m_{\chi}$, one would need to eventually include the decays of $\mathcal{S}$ to a pair of low-lying KK fermions, which will make $\mathcal{S}$ much wider of what is sensible in a perturbative theory.

In addition, we display for comparison the contribution due to diagrams mediated by the first KK graviton, which are also irreducible in models with WEDs (see e.g. \cite{Arcadi:2019lka,Lee:2013bua} for useful expressions). We can see that, for the chosen values of $\mkk$ and $k/m_{\rm Pl}$, corresponding to $\mkk=5$ TeV and $\Lambda_{\pi}=m_{\rm Pl}e^{-k\pi r}=40$ TeV, the contribution of the odd scalar resonance $\mathcal{S}$ dominates over the KK graviton one. In particular, this happens for all values of $m_{\chi}$, with the exception of the small region where the coupling $y_{\chi\mathcal{S}}$ goes to zero. The relative importance of each contribution and the location of the graviton peak can be changed by modifying the ratio $\Lambda_{\pi}/\mkk$ and/or by including brane kinetic terms \cite{Davoudiasl:2003zt}. We will not explore such possibilities, being our aim here to show that the scalar contribution can naturally be the leading one, as one can readily see from the figure. In addition to the KK-graviton contribution, one also expects a contribution to the coannihilation cross section arising from the exchange of the radion. This contribution is rather model dependent, since the radion mass is subject to the specifics of the stabilization mechanism. A natural expectation is that the radion is  much lighter that the first KK graviton. This case was considered e.g.\ in~\cite{ Folgado:2020vjb}, where the authors considered a light radion interacting with IR-localized matter and found the radion contribution to be mostly irrelevant. A similar result is expected here, for a light radion not mixing with the other bulk scalars. The interesting case where the stabilizing scalar mixes with both the Higgs and the $\mathbb{Z}_2$-odd scalar would  require a fairly  extensive case study, which is beyond of the scope of this paper. 

Finally, we also show the values of the velocity averaged cross section for which the observed DM relic abundance is reproduced, both in the usual scenario and in the case of an early period of matter domination. In particular, we show in dashed black the values of $\langle \sigma v \rangle$ for which a value of $\Omega_{\chi} h^2=0.12$ is reproduced, in the case of a regular freeze-out mechanism, and in the scenario of matter domination in gray, for $\tau=0.99$, $T_{\star}=10^5$ GeV and two values of $T_{\rm RH}$, 1 and $10^2$ GeV, respectively. The lines in dot-dashed gray correspond to regions where $x_f<3$, where the DM is expected to decouple relativistically and the current treatment loses validity, see \cite{Hamdan:2017psw,Chanda:2019xyl} for more details. We can see that the observed relic abundance can be reproduced in the case of matter domination for masses $ m_{\chi}\sim 8-10$ TeV. In the usual case of radiation domination, $\langle \sigma v\rangle $   can be a non-negligible fraction of the one which is required to reproduce the observed relic abundance for $m_{\chi}\sim 15$ TeV, which is in the ballpark of the naturally expected fermion masses.

	\subsection{Direct detection}\label{sec:xenon}

Direct detection experiments can also set very important constraints on the parameter space in scalar-mediated models of DM. Indeed, they constraint all the parameter space in the case of Higgs-mediated DM, with the exception of a small region around the Higgs resonance, see e.g.~\cite{Aghanim:2018eyx}. We study here the constraints from direct detection experiments in our model. In particular we will compare our predictions with results from Xenon1T \cite{DiGangi:2018xht, Aprile:2018dbl}. We are interested in the spin-independent cross section
\begin{equation}
	\sigma_{\chi N} \approx \dfrac{4}{\pi} \mu^2_{\chi N} \left[ Z f_p + (A-Z) f_n \right]^2 \simeq \dfrac{4}{\pi} \mu^2_{\chi N} A^2 f_n^2 ,
\end{equation}
with $Z$ and $A$ the atomic number and atomic mass of the target nucleus, respectively, and $\mu_{\chi N}$ the reduced mass of the DM and nucleus system \cite{Kim:2006af, Kanemura:2010sh, Arcadi:2019lka}. In order to compute such cross section we use following effective Lagrangian
\begin{equation}\mathcal{L}_\text{eff} = f_p (\bar \chi \chi)(\bar p p) + f_n (\bar \chi \chi)(\bar n n ).
\end{equation}

The terms $f_p$ and $f_n$ are effective coupling constants and can be written as 
\begin{equation}
\frac{f_{p,n}}{m_{p,n}} = \sum_{q=u,d,s} f_{Tq}^{(p,n)} \frac{\alpha_q}{m_q}
 + \frac{2}{27} f_{Tg}^{(p,n)} \sum_{q=c,b,t} \frac{\alpha_q}{m_q} ,
\end{equation}
where $\alpha_q$ stands for the effective four-fermion interaction vertex, obtained by considering the scalar t–channel exchange. In our model $\alpha_q$ has the following form
\begin{equation}
	\alpha_q = y_{\chi \mathcal{S}}\left\lbrace \dfrac{y_{q\mathcal{S}}}{m_\mathcal{S}^2} + \dfrac{y_{q h} \sin \theta_{h\mathcal{S}} }{m_h^2} \right\rbrace . 
\label{eq:alpha_direct}
\end{equation}

\begin{figure*}[t!]
\begin{center}
\hspace*{-0.25cm}
\includegraphics[scale = 0.55]{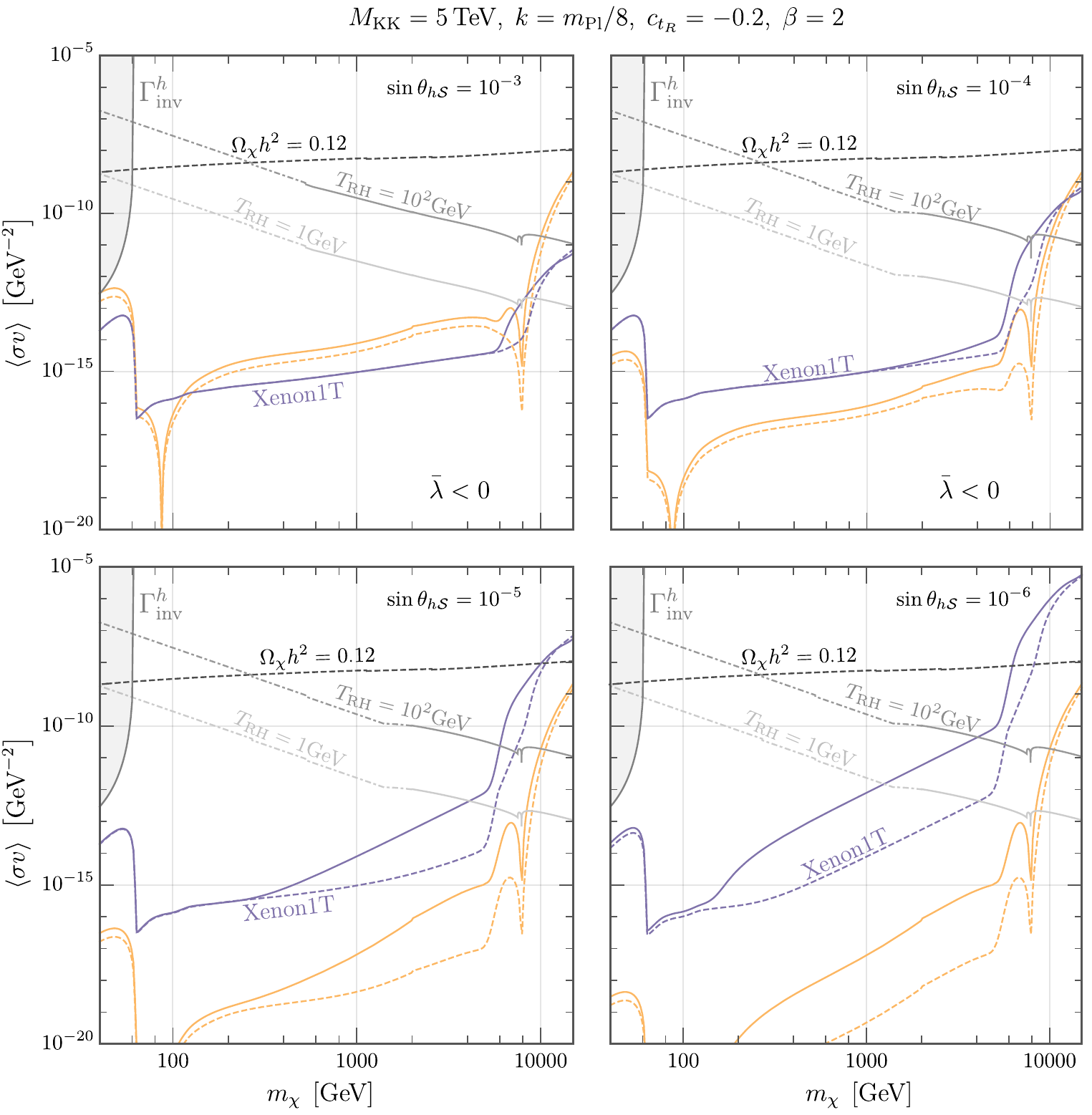}
\vspace*{-0.25cm}
	\caption{Velocity averaged coannihilation cross section at the freeze-out temperature for different values of the mixing between the odd scalar and the Higgs boson, $\sin \theta_{h\mathcal{S}} = \lbrace 10^{-3}, 10^{-4}, 10^{-5}, 10^{-6} \rbrace $, from top left to bottom right. The first two cases correspond to negative values of $\bl$. We have set $N_{\chi}=1$, $\mkk=5$~TeV and $k=m_{\rm Pl}/8$. We show in yellow the predictions for two different benchmarks with different values of $y_{\ast}$ and $\lambda_{S}/r$. In both cases, we have fixed $c_{t_R}=-0.2$ and $\beta=2$. We show the constraints coming from the Higgs invisible decay width in gray and the limits from Xenon1T in purple. We show the velocity averaged cross section reproducing the relic density experimental value from Planck in dashed black, and the equivalent for a matter dominated freeze out in gray, for two different values of $T_{\rm RH}$, where we used $T_\star = 10^5$ GeV and $\tau=0.99$. For these lines the section in dot-dashed gray corresponds to predictions for which $x_f < 3$, and therefore in this regime the DM decouples relativistically \cite{Hamdan:2017psw,Chanda:2019xyl}.} \label{fig:cross_total}
\end{center}
\end{figure*}

Finally, $f_{Tg}^{(p,n)}$ is defined as
\begin{equation}
f_{Tg}^{(p,n)} = 1-\sum_{q=u,d,s} f_{Tq}^{(p,n)},
\end{equation}
and the values for $f^q_p$ and $f^q_n$ are \cite{Hoferichter:2015dsa, Arcadi:2019lka}
\begin{equation}
\hspace*{-0.2cm}
\begin{aligned}
f^u_p = (20.8 \pm 1.5) \cdot 10^{-3}, && f^d_p = (41.1 \pm 2.8) \cdot 10^{-3}, \\
f^u_n = (18.9 \pm 1.4) \cdot 10^{-3}, && f^d_n = (45.1 \pm 2.7) \cdot 10^{-3}, \\
f^s_p = f^s_n = 0.043 \pm 0.011.
\end{aligned}
\end{equation}

One can compare the contribution of each scalar to the direct detection cross section by computing the ratio between the terms appearing in equation~\eqref{eq:alpha_direct}. We find that the channel mediated by the Higgs boson is dominant provided that 
\begin{equation}
	\sin \theta_{h\mathcal{S}} > \dfrac{m_h^2}{m_\mathcal{S}^2} \dfrac{y_{q\mathcal{S}}}{y_{q h}} \sim 10^{-7},
\end{equation}
i.e.~we expect the Higgs mediated interaction to be the leading contribution for $\sin \theta_{h\mathcal{S}} > 10^{-7}$. This tells us in particular that we can relax the constraints coming from direct detection by making the mixing smaller, while keeping the same coupling $y_{\chi \mathcal{S}}$ to the DM fermions. However, this is only possible up to the point when the odd scalar contribution becomes dominant, 
\begin{equation}
	(\alpha_{q})_{\min}\approx \frac{y_{\chi \mathcal{S}} y_{q\mathcal{S}}}{m_{\mathcal{S}}^2}.
\end{equation}

We show in figure~\ref{fig:cross_total} the constraints coming from direct detection and invisible Higgs decays for the velocity averaged coannihilation cross section $\langle \sigma v\rangle$ as a function of $m_{\chi}$. We used $N_{\chi}=1$, $\mkk=5$ TeV, $k=m_{\rm Pl}/8$ and different mixing values between the odd scalar and the Higgs boson, $\sin \theta_{h\mathcal{S}} = \lbrace 10^{-3}, 10^{-4}, 10^{-5}, 10^{-6} \rbrace $, from top left to bottom right. The first two mixing angles can only be achieved for $\bl<0$, whereas the last two can be obtained for positive and negative values of $\bl$. We display in each figure two different benchmarks, corresponding to the choices $y_{\ast}=3$ (solid line) and $y_{\ast}=1.5$ (dashed line) for the third generation quarks $t$ and $b$ (as before, light generations have identical bulk mass parameters in absolute value and different values of $y_{\ast}$, starting with $y_{\ast}^s=1/2$). In both cases, we have set $\beta=2$ and $c_{t_R}=-0.2$, while $\lambda_S/r$ has been chosen in such a way that $\Gamma_{\mathcal{S}}/m_{\mathcal{S}}\approx 0.7$. More specifically, we have taken $\lambda_S/r=120$ and $\lambda_S/r=65$, for $y_{\ast}=3$ and $y_{\ast}=1.5$, respectively. Since the width is mostly given by the decay of $\mathcal{S}$ into a third generation quark and its first KK resonance, such assignment ensures that the overall coupling of the odd scalar field to the visible sector is roughly the same in both cases. However, the smaller value of $y_{\ast}$ in the benchmark $\{y_{\ast}=1.5,~\lambda_S/r=65\}$ leads to a more IR-localized  third-generation left-handed doublet $q_L^3$ and to a much larger coupling of $\mathcal{S}$ to $\bar{b}_L b_R$ and $\bar{q}_L^3$ plus its first KK resonance, even with a smaller value of $\lambda_S/r$. At the end of the day, however, the solid lines are above the dashed ones for most values of $m_{\chi}$, since the DM coupling $y_{\chi \mathcal{S}}$ is smaller by a factor $\sqrt{120/65}\sim 1.4$, which makes up for the small differences existing among the couplings to the visible sector. The differences between both benchmarks are magnified once the purely $\mathcal{S}$-mediated channel, corresponding to the right column of figure~\ref{fig:diagramsxsec}, is the most dominant one. This happens in particular for large DM masses and/or  small values of $\sin\theta_{h\mathcal{S}}$, as one can readily see by comparing the different panels in  figure~\ref{fig:cross_total}.

The gray region shows the area excluded by the LHC experimental limits on the Higgs invisible decay width, and in purple we show the Xenon1T constraints. The latter are found by plotting the velocity averaged coannihilation cross section obtained after rescaling $y_{\chi \mathcal{S}}$ such that $\sigma_{\chi N}$ saturates the Xenon1T experimental bound, $\langle \sigma v\rangle_{\rm Xenon1T}$. For the values of $\sin\theta_{h\mathcal{S}}$ shown in this figure, the leading contribution to the DM-nucleon cross section is by far the one arising from the $t$-channel exchange of a Higgs boson, with the exception of the last case where $\sin\theta_{h\mathcal{S}}=10^{-6}$ and the $\mathcal{S}$ contribution, while still subleading, starts to be relevant. 
This explains why the Xenon1T bound for the $\{y_{\ast}=3,~\lambda_S/r=120\}$ benchmark is weaker than the limit obtained for $\{y_{\ast}=1.5,~\lambda_S/r=65\}$, whenever the coannihilation cross section is dominated by the $\mathcal{S}$ contribution. Indeed, in the former case, the couplings of $\mathcal{S}$ to the visible sector are slightly larger. This leads to a larger value of $\langle \sigma v\rangle_{\rm Xenon1T}$ after rescaling $y_{\chi\mathcal{S}}$ and to a weaker bound from direct detection. When $\langle \sigma v\rangle $ is dominated by the Higgs exchange, direct detection bounds become indistinguishable for both benchmarks, since the Higgs couplings to the SM quarks are mostly fixed and SM-like.

We also show the velocity averaged cross section reproducing the observed relic density both in the usual freeze-out scenario (dashed black) and in the case of an early period of matter domination, for values of $T_{\rm RH} = 10^2$ GeV (dark gray) and 1 GeV (light gray). For both gray lines, we used $T_\star = 10^5$ GeV and $\tau = 0.99$. Similarly to figure~\ref{fig:cross_products}, lines in dot-dashed gray correspond to regions where $x_f < 3$ and the DM is expected to decouple relativistically. We can see that for $\sin\theta_{h\mathcal{S}}=10^{-3}$, one can not explain the observed relic abundance without exceeding the bounds from Xenon1T. However, this is not the case in the matter dominated scenario with $T_{\rm RH}=1$~GeV, where the required coannihilation cross section to explain the DM relic abundance does not exceed the Xenon1T bound for $y_{\ast}=3$. In the case of $y_{\ast}=1.5$, the required cross section is excluded by the Xenon1T bound. In the case of $\sin\theta_{h\mathcal{S}}=10^{-4}$ we can reproduce the correct amount of DM for both values of $T_{\rm RH}$, in the scenario of matter domination, being the values of $\langle \sigma v\rangle $ corresponding to the top of the resonant peak excluded by direct detection bounds. For even smaller values of $\sin\theta_{h\mathcal{S}}$ like $10^{-5}$ or $10^{-6}$, the data from Xenon1T never constrains the predictions for the coannihilation cross section obtained in both benchmarks, since the Higgs coupling to DM $y_{\chi h}$ becomes too small. Therefore, by assuming an early period of matter domination, we are able to explain the observed DM relic abundance for moderately small values of $\sin\theta_{h\mathcal{S}}$ without conflicting current direct detection experiments. Even in the case of radiation domination, we can get to values of $\langle \sigma v\rangle$ relatively close to the ballpark of what is needed, expecting $\mathcal{S}$ to be a non-negligible fraction of the required coannihilation cross section, even though additional mediators accounting for most of the coannihilation are certainly needed.

\section{Summary}

We have demonstrated that the addition of a $\mathbb{Z}_2$-odd scalar field developing a VEV in extra-dimensional models can not only account for the origin of the 5D fermion masses, but also provide a unique window into any 5D fermionic dark sector. Indeed, since such a scalar field generates dynamically fermion bulk masses through Yukawa-like interactions with the different 5D fermions, it will also irrevocably connect  the SM with  any possible dark sector featuring bulk fermions. Moreover, in realistic models the Higgs scalar field propagates into the bulk of the WED, and thus a mixing with the new scalar field is unavoidable. In this work, we have studied in detail the phenomenological consequences of such a portal, showing that the lightest KK dark fermion is stable and can coannihilate efficiently thanks to the mediation of the odd-scalar resonances as well as the Higgs boson. Indeed, we have demonstrated that it is possible to reproduce the observed DM relic abundance for an $\mathcal{O}(10)$~TeV KK dark fermion assuming that freeze-out occurs during an early period of matter domination, without conflicting with current data from direct-detection experiments. Even in the regular case of a radiation dominated freeze-out, this irreducible contribution to the coannihilation cross-section can account for a non-negligible part of the required value when the DM mass is $\sim 15$~TeV. We have also shown that these scalar contributions to the coannihilation cross section can be more important than those arising from the exchange of KK gravitons. The bounds arising from direct detection are only relevant when the parameter $\sin\theta_{h\mathcal{S}}$ controlling the mixing between the SM-like Higgs boson and the first KK resonance $\mathcal{S}$ of the $\mathbb{Z}_2$-odd scalar field is $\gtrsim 10^{-4}$. For smaller values, the contribution to the direct-detection cross section given by $t$-channel Higgs exchange becomes less and less important, to the point of becoming of the same order as the one from the $t$-channel exchange of the $\mathcal{S}$ resonance, which is beyond the reach of current direct detection experiments.

We have also studied the impact of the scalar mixing on precision measurements of Higgs couplings. In particular, we have computed the modifications of the Higgs couplings to electroweak gauge bosons and the bottom quark  as a consequence of the mixing between the SM-like Higgs boson and the first KK resonances of both bulk fields, $\mathcal{H}$ and $\mathcal{S}$. We have demonstrated that planned future colliders  could probe the induced modifications on the $b$-quark Yukawa in the case where $\beta\lesssim 4$, values for which the Higgs boson has a strong presence into the bulk. We have also studied the constraints on the Higgs effective Yukawa coupling to DM when its mass is light enough to allow for the Higgs boson to decay into a pair of DM particles. We conclude that the effective Yukawa coupling to the dark fermions $y_{\chi h}\lesssim 0.02/\sqrt{N_{\chi}}$, with $N_{\chi}$ being the multiplicity of the 5D dark fermion.

In summary, we have shown that models with a WED naturally feature a compelling explanation for the observed relic abundance of DM, consisting of an $\mathcal{O}(10)$ TeV fermionic WIMP coupled to the SM by a heavy scalar mediator $\mathcal{S}$ with mass $m_{\mathcal{S}}\sim 30$~TeV. All this is possible without conflicting with current data from colliders, flavor experiments and cosmology and while still providing natural solutions to the hierarchy problem and the flavor puzzle, which are arguably two of the most important theoretical problems in particle physics. 

\begin{acknowledgments}
  We would like to thank Miki Chala and Aqeel Ahmed for useful comments on the manuscript, as well as Lukas Mittnacht and Eric Madge for  helpful discussions.  M.N.~thanks Gino Isidori, the particle theory group at Zurich University and the Pauli Center for hospitality during a sabbatical stay. A.C.\ acknowledges funding from the European Union’s Horizon 2020 research and innovation programme under the Marie Skłodowska-Curie grant agreement No.~754446 and UGR Research and Knowledge Transfer Found – Athenea3i. The research of J.C.\ and M.N.\ is supported by the Cluster of Excellence {\em Precision Physics, Fundamental Interactions and Structure of Matter\/} (PRISMA${}^+$ -- EXC~2118/1) within the German Excellence Strategy (project ID 39083149) and under grant 05H18UMCA1 of the German Federal Ministry for Education and Research (BMBF).\end{acknowledgments}


\appendix


\bigskip\bigskip

\section{Fermion equations of motion} \label{app:fermions}

The Yukawa interaction between the odd bulk scalar $S_1$ and a $\bar \Psi_a \Psi_A$ pair, corresponding to a fermion zero mode (a) and a first KK mode (A), comes from the term generating the different fermion bulk masses
\begin{equation}
S \supset - \int d^5x \sqrt{g}\, \mathcal{Y}_a \bar \Psi_a \Psi_A \Sigma .
\end{equation}
In this case, the EOM for a fermion field with a 5D bulk mass generated dynamically reads
\begin{equation}
\left[\pm t \partial_t - c_a \, v_S(t) \right] f_{0,a}^{L,R} (t) = 0,
\end{equation}
for the zero mode and 
\begin{equation}
\begin{aligned}
\left[ t^2 \partial_t^2 \right. & + x_n^2 t^2 \mp c_a \, t \, v_S'(t) \\
& \left. + c_a \, v_S(t) \left(\pm 1 - c_a \, v_S(t) \right) \right] f_{n,A}^{L,R}(t) = 0, \label{eq:kkferm}
\end{aligned}
\end{equation}
for the $n$-th KK mode \cite{Ahmed:2019zxm}, where $c_a$ is the usual dimensionless 5D mass. In the scenario at hand, it is defined by \cite{Ahmed:2019zxm}
\begin{equation}
c_a \equiv \mathcal{Y}_a \sqrt{\dfrac{6}{\lambda_S}} \dfrac{\mu_S}{k}.
	\label{eq:cobrakai}
\end{equation}
 Here, we have used the following KK decomposition for the fermions
\begin{equation}
\Psi = \sum_{n=0} \Psi_n(x) \left(\dfrac{t}{\epsilon}\right)^2 \sqrt{\mkk} \, f_n(t) ,
\end{equation}
satisfying
\begin{equation}
2 \int_\epsilon^1 f_m^{A*}(t) f_n^{B}(t) = \delta_{mn}.
\end{equation}
The 5D Yukawa interaction leads to the effective 4D vertex $S_1 \bar \Psi_a \Psi_A$, with effective Yukawa
\begin{equation}
y_{a A \mathcal{S}} = 2 \, c_a \sqrt{\dfrac{\lambda_S}{r}} \dfrac{k}{\mu_S} \int_\epsilon^1 dt f_a(t) f_A(t) \chi_1^S(t).
	\label{eq:oddyuk}
\end{equation}

The coupling to a VL fermion $\chi$, corresponding to the first KK resonance of some 5D fermion, is given by
\begin{equation}
	y_{\chi\mathcal{S}} = 2 \, c_a \sqrt{\dfrac{\lambda_S}{r}} \dfrac{k}{\mu_S} \int_\epsilon^1 dt f_{1,\chi}^{L}(t) f_{1,\chi}^{R}(t) \chi_1^S(t).
	\label{eq:oddyukdm}
\end{equation}

\section{Cross section expressions}

The different coannihilation cross sections of $\bar{\chi}\chi$ into a pair of SM particles are given by (see e.g.~\cite{Chanda:2019xyl})

\begin{widetext}
\begin{equation}
\begin{aligned}
	\sigma(\bar \chi \chi \to \bar f f ) & = \dfrac{N_{\! f} \, y_{\chi\mathcal{S}}^2}{16 \pi} s \left( 1 - \dfrac{4 m_f^2}{s} \right)^{3/2} \left( 1 - \dfrac{4 m_\chi^2}{s} \right)^{1/2} \left[ \dfrac{y_{f h}^2 \sin^2 \theta_{h\mathcal{S}}}{\left(s-m_h^2 \right)^2} + \dfrac{y_{f\mathcal{S}}^2}{\left(s-m_\mathcal{S}^2 \right)^2} + \dfrac{2 \, y_{f h} y_{f\mathcal{S}} \sin \theta_{h\mathcal{S}}}{\left(s-m_\mathcal{S}^2 \right)\left(s-m_h^2 \right)} \right], \\
	\sigma(\bar \chi \chi \to hh ) & = \dfrac{y_{\chi\mathcal{S}}^2}{32 \pi} \left( 1 - \dfrac{4 m_h^2}{s} \right)^{1/2} \left( 1 - \dfrac{4 m_\chi^2}{s} \right)^{1/2} \left[ \dfrac{9 \sin^2 \theta_{h\mathcal{S}} m_h^4}{v_4^2 \left(s-m_h^2 \right)^2} + \dfrac{\sin^2 \theta_{h\mathcal{S}} x_{S_1}^4 \mkk^4}{v_4^2 \left(s-m_\mathcal{S}^2 \right)^2} + \dfrac{6 \,m_h^2 \sin^2 \theta_{h\mathcal{S}} x_{S_1}^2 \mkk^2 }{v_4^2 \left(s-m_\mathcal{S}^2 \right)\left(s-m_h^2 \right)} \right], \\
	\sigma(\bar \chi \chi \to VV ) & = \dfrac{\delta_V y_{\chi\mathcal{S}}^2 \sin^2 \theta_{h\mathcal{S}}}{8 \pi} \left( 1 - \dfrac{4 m_V^2}{s} \right)^{1/2} \left( 1 - \dfrac{4 m_\chi^2}{s} \right)^{1/2} \dfrac{m_V^4}{v_4^2\left(s-m_h^2 \right)^2} \left[ 2 + \dfrac{(s-2m_V^2)^2}{4m_V^4} \right], \\
	\sigma(\bar \chi \chi \to \bar Q q ) & = \dfrac{N_{\! Q} \, y_{\chi\mathcal{S}}^2 y_{Qq}^2}{16 \pi} \dfrac{s}{\left(s-m_\mathcal{S}^2 \right)^2} \left( 1 - \dfrac{m_Q^2}{s} \right)^{3/2} \left( 1 - \dfrac{4 m_\chi^2}{s} \right)^{1/2} , \\
\end{aligned}
\end{equation}
\end{widetext}
with $\delta_V = 1, 1/2$ for $V = W^\pm, Z$.

\section{Matter dominated freeze-out} \label{app_mdf}

We review here the relevant formulae for the calculation of the current DM relic abundance after freeze-out during an early period of matter domination (see \cite{Hamdan:2017psw,Chanda:2019xyl} for more details). 

We assume the presence of a long-lived heavy scalar field $\phi$, localized on the UV brane, which starts behaving like matter at a critical temperature $T_{\star}\sim m_{\phi}$, which we assumme to be much larger than $\mkk$, i.e., $T_{\star}\gg \mkk$. If $\phi$ is long-lived enough, its contribution to the energy density $\rho_{\phi}$ will grow until ultimately monopolize the total energy density regardless of its initial contribution at $T_{\star}$, given by $(1 - \tau)$, with $\tau\in [0, 1]$
\begin{equation}
	\tau=\left.\frac{\rho_R+\rho_\chi}{\rho_{R}+\rho_{\chi}+\rho_{\phi}}\right|_{T=T_{\star}},
\end{equation}
and $\rho_{\chi}$, $\rho_{R}$, the energy density of DM and the visible sector of the extra-dimensional theory. 

Using the Friedmann equation 
\begin{align}
	\begin{aligned}
		H^2=\frac{1}{3 m_{\rm Pl}^2}\left[\rho_R+\rho_{\chi}+\rho_{\phi}\right] \\
	\end{aligned}
\end{align}
and defining $H_{\star}=H(T_{\star})$ we obtain
\begin{widetext}
\begin{equation}
	H^2=H_{\star}^2\left[\frac{g_{\ast} \tau}{g_{\ast}+g_{\chi}}\left(\frac{a_{\star}}{a}\right)^4+(1-\tau)\left(\frac{a_{\star}}{a}\right)^3+\frac{g_{\chi} \tau}{g_{\ast}+g_{\chi}}\left(\frac{a_{\star}}{a}\right)^4\right],
\end{equation}
\end{widetext}
where $g_{\ast}$ is the effective number of relativistic degrees of freedom of the visible sector of the extra-dimensional theory. Assuming that the entropy is conserved in this sector and taking into account that $g_{\ast}\gg g_{\chi}$, we can write
\begin{equation}
	H=H_{\star}\sqrt{1-\tau}\left(\frac{x_{\star}}{x}\right)^{3/2}\left[\frac{r}{1-\tau}\left(\frac{x_{\star}}{x}\right)+1\right]^{1/2},
\end{equation}
where we have defined $x=m_{\chi}/T$ and $x_{\star}=x(T_{\star})$. 

The annihilation rate $\Gamma_{\rm ann}$ can be written as 
\begin{equation}
	\Gamma_{\rm ann}=\frac{g_{\chi} m_{\chi}^3}{(2\pi x)^{3/2}}e^{-x}\langle \sigma v\rangle 
\end{equation}
where $\langle \sigma v\rangle$ is the velocity-averaged coannihilation cross section. In our scenario of scalar mediated fermionic DM, after Taylor expanding $\left( \sigma v\right)$ with respect to $v_r^2$, we can write 
\begin{equation}
\left( \sigma v \right) \simeq a + b v_r^2 = b v_r^2 ,
\end{equation}
since $a = 0$ as we have seen in the previous section. Therefore, one can write the velocity-averaged coannihilation cross section as
\begin{equation}
\left\langle \sigma v \right\rangle \simeq \dfrac{6b}{x},
\end{equation}
which leads to
\begin{align}
	\Gamma_{\rm ann}=\frac{g_{\chi} m_{\chi}^3}{(2\pi)^{3/2}}e^{-x}\frac{6b}{x^{5/2}} .
\end{align}
We can define the freeze-out temperature by asking $H(x_f)=\Gamma_{\rm ann}(x_f)$, with $x_f=m_{\chi}/T_{f}$. This leads to
\begin{widetext}
\begin{equation}
\begin{aligned}
	x_f = \log \left(\frac{g_{\chi}}{(2\pi)^{3/2}}\frac{m_{\chi}^3 6b}{H_\star x_\star^{3/2}} (1-\tau)^{-1/2} \left( x_f^{2} + \dfrac{r x_\star}{1-\tau} x_f \right)^{1/2} \right) .
\end{aligned}
\end{equation}
\end{widetext}

The yield $Y=n_{\chi}/s$, with $s$ the entropy density, is given at freeze-out by
\begin{equation}
Y_f = \left(\lambda \int^\infty_{x_f} dx \left( 1+ \dfrac{\tau}{1-\tau}\dfrac{x_\star}{x} \right)^{-1/2} x^{-7/2} \right)^{-1} ,
\end{equation}
where
\begin{equation}
	\lambda = \dfrac{2 \pi^2 g_{\ast S} m_\chi^3 6 b}{45 H_\star x_\star^{3/2}},
\end{equation}
where $g_{\ast S}$ is the number of effective entropic degrees of freedom. More explicitly, it reads
\begin{widetext}
\begin{equation}
\begin{aligned}
	\frac{1}{Y_f}\approx \frac{\pi^2 g_{\ast S}m_{\chi}^3 \langle \sigma v\rangle }{90 H_{\star}x_{\star}^4}\left(\frac{1-\tau}{\tau}\right)^{5/2} & \Bigg(3 x_f \sinh^{-1}\left(\sqrt{\frac{x_{\star} \tau}{x_f(1-\tau)}}\right) \\ 
	& + \sqrt{\dfrac{x_{\star}}{x_f}} \left(\dfrac{\tau}{1-\tau}\right)^{3/2} \sqrt{1+\frac{x_{\star}\tau}{x_f(1-\tau)}}\left[ 2x_{\star} - 3x_f\frac{(1-\tau)}{\tau} \right]\Bigg),
\end{aligned}
\end{equation}
\end{widetext}
where we have approximated $\langle \sigma v \rangle \approx 6b/x_f$.
Finally, the prediction for the DM relic abundance, assuming a matter dominated universe during freeze-out, reads
\begin{equation}
\Omega_\chi h^2 = \zeta \dfrac{s_0 m_\chi Y_f}{\rho_{\text{critical}}} ,
\end{equation}
where $\rho_{\text{critical}} = 8.13\cdot 10^{-47}$ GeV$^4$, $s_0 \simeq 2.29 \cdot 10^{-38}$ GeV$^3$. In the above equation, $\zeta$ parametrizes the dilution of the DM abundance after freeze-out due to the decays of $\phi$ and the subsequent entropy injection 
\begin{align}
	\zeta = \frac{s_{\rm before}}{s_{\rm after}}=\frac{\Omega_{\chi}}{\Omega_{\chi}^f},
\end{align}
which can be expressed with good approximation as follows 
\begin{equation}
	\zeta \approx \dfrac{45}{4 \pi^3} \dfrac{1}{(1-\tau)g_\ast} \dfrac{T_{\text{RH}}}{T_{\star}}.
\end{equation}


\bibliographystyle{JHEP}
\bibliography{refs.bib}


\end{document}